# Effects of $CO_2$ flushing on crystal textures and compositions: experimental evidence from recent K-trachybasalts erupted at Mt. Etna

Marisa Giuffrida[1], Francois Holtz[2], Francesco Vetere[3], Marco Viccaro[1,4]

[1] *Università di Catania, Dipartimento di Scienze Biologiche, Geologiche e Ambientali – Sezione di Scienze della Terra, Corso Italia 57, I-95129, Catania, Italy*

[2] *Leibniz Universität Hannover, Institut of Mineralogy, Callinstr. 3, D-30167 Hannover*

[3] *Università di Perugia, Dipartimento di Fisica e Geologia, Piazza Università 1, Perugia, Italy*

[4] *Istituto Nazionale di Geofisica e Vulcanologia, Osservatorio Etneo, Piazza Roma 2, I-95125, Catania Italy*

**Abstract**

Changes in magmatic assemblages and crystal stability as a response of $CO_2$-flushing in basaltic systems have been never directly addressed experimentally, making the role of $CO_2$ in magma dynamics still controversial and object of scientific debate. We conducted a ~~A~~ series of experiments to understand the response of magmas from Etna volcano to $CO_2$ flushing. We performed a first experiment at 300 MPa to synthesize a starting material composed of crystals of some hundreds of μm and melt pools. This material is representative of an initial magmatic assemblage composed of plagioclase, clinopyroxene and a water-undersaturated melt with 1.6 wt.% $H_2O$. In a second step, the initial assemblage was equilibrated at 300 and 100 MPa with fluids having different $XCO_2^{fl}$ ($CO_2/(H_2O + CO_2)$). At low $XCO_2^{fl}$ (< 0.2 to 0.4), plagioclase is completely dissolved and clinopyroxene show dissolution textures. For relatively high $XCO_2^{fl}$ (0.9 at 300 MPa), the flushing of a $CO_2$-rich fluid phase leads to an increase of the amount of clinopyroxene and a decrease of the abundance of plagioclase at 300 MPa. This decrease of plagioclase proportion is associated with a change in An content.

Our experiments demonstrate that flushing basaltic systems with fluids may drastically affect crystal textures and phase equilibria depending on the amount of $H_2O$-$CO_2$ in the fluid phase. Since texture and crystal proportions are among the most important parameters governing the rheology of magmas, fluid flushing will also influence magma ascent to the Earth's surface. The experimental results open new perspectives to decipher the textural and compositional record of minerals observed in volcanic rocks from Mt. Etna, and at the same time offer the basis for interpreting the information preserved in minerals from other basaltic volcanoes erupting magmas enriched in $CO_2$

Keywords: Mt. Etna, crystallization experiments, $CO_2$ flushing, crystal textures.

## 1. Introduction

Processes of gas escape and changing concentration of volatiles in natural magmas provide the driving force for volcanic eruptions, as volatiles control the physical melt properties and phase equilibria, with far-reaching implications for magma ascent dynamics and eruptive styles (e.g., Sparks and Pinkerton 1978; Papale and Dobran 1994; Dingwell 1998; Papale 1998; Blundy and Cashman 2001; Sparks et al. 2003; Cashman 2004; Webster et al. 2015). Magmatic volatile composition is typically dominated by $H_2O$ and $CO_2$. These two components behave quite differently in the melt phase during ascent, due to their strong pressure-dependent and contrasting solubilities (e.g., Brey 1976; Papale 1997; Webster et al. 1999; Newman and Lowenstern 2002; Vetere et al. 2011; Iacono-Marziano et al. 2012; Witham et al. 2012; Vetere et al. 2014; Shishkina et al. 2014a; 2014b). While the central influence of water on volcanic processes has been recognized since decades (e.g., Tuttle and Bowen 1958), the role of $CO_2$ has only recently been considered to play a central role in magmatic and volcanic dynamics (e.g., Wilson et al. 1980; Spera 1981; 1984; Papale and Polacci 1999; Aiuppa et al. 2007; Blundy et al. 2010; Nicotra and Viccaro 2012). In particular, understanding the behavior of mixed $H_2O$-$CO_2$ volatile components in experimental studies has become one of the key to unravel the effects of ascent rates on the kinetics of magma vesiculation and crystallization, along with their implications on eruption mechanisms (e.g., Mourtada-Bonnefoi and Laporte 2002; Cichy et al. 2011; Pichavant et al. 2013; Riker et al. 2015; Fiege et al. 2015). In spite of the increasing interest in reproducing crystallization paths of multicomponent volatile-bearing systems, our knowledge on the concomitant effect of mixed $H_2O+CO_2$ fluids on phase relations in alkali-rich basaltic systems is restricted to a more limited number of experimental data (e.g., Freise et al. 2009; Pichavant et al. 2013; 2014; Vetere et al. 2015a). These studies primarily focused on the evolution of phase relations and melt composition across a range of physical and chemical conditions (i.e. P, T, $H_2O$ content), providing information about crystal changes in terms of abundances and size. Previous studies, however, did not directly address changes in magmatic assemblages and crystal stability as a response of fluid flushing which causes local variations in the $CO_2$ proportion of the magmatic system. The interpretation of textural features of crystals has been, so far, conducted through experimental reproduction of processes of crystal dissolution/resorption or their growth in presence of a pure-$H_2O$ magmatic fluid phase (Logfren 1980; Nelson and Montana 1989; Tsuchiyama 1985; Tsuchiyama and Takahashi 1983; Kawamoto 1992; Kirkpatrick 1981; Hummer and Rutherford 2002). Thus, the influence of volatile components in addition to water on textural and compositional changes in alkali basalts has been generally neglected. This means that mineral reactions in systems undergoing rapid changes of ($CO_2$-bearing) fluid compositions is still not fully understood, precluding in some cases a proper interpretation of natural samples and of related physical processes during magma ascent.

$CO_2$ flushing from magmas exsolving gas at depth has been found to play a major role in producing highly explosive eruptions at many basaltic systems such as Etna and Stromboli (cf. Aiuppa et al. 2010a; 2010b; Aiuppa et al., 2016; Allard 2010; Pichavant at al. 2013; Ferlito et al. 2014). In this context, Mt. Etna represents one of the most intriguing cases. The complex plumbing system of Mt. Etna is persistently fed by primitive, volatile-rich magmas that mix with mafic, degassed magmas stored at lower pressure (Armienti et al. 2007; 2013; Ferlito et al. 2008; Viccaro et al. 2010; 2014; 2015; Nicotra and Viccaro 2012a; Corsaro et al. 2013; Mollo et al. 2015). Moreover, published data on primitive glass inclusions in olivine crystals indicate that magmas of Mt. Etna can exsolve a gas phase, rich in $CO_2$, at pressure higher than 250 MPa (Metrich et al., 2004; Spilliert et al., 2006; Collins et al. 2009), which continuously flush the overlying magma reservoirs, acting as a recurrent trigger mechanism for paroxysmal events (Aiuppa et al., 2007, 2016; Nicotra and Viccaro 2012; Ferlito et al. 2014). This means that a major part of the crystallization occurs in regions of the plumbing system where $CO_2$ is already exsolved, but in which $H_2O$ activity can change very abruptly as a result of local fluxes of $CO_2$-rich fluids in magma reservoirs (cf. Aiuppa et al. 2010b; Patanè et al. 2013).

In this study, we tested experimentally the effect of $CO_2$ flushing in a partially crystallized K-trachybasaltic magma representative of eruptive products from the recent activity of Mt. Etna. We demonstrate that the flushing of $CO_2$ results in variations of mineral stability and phase relations, which are consistent with observations made in rock samples. The experimental results obtained at 300 and 100 MPa are applied to test the role of $CO_2$ on mineral textures and melt chemistry during storage in the intermediate (9-6 Km b.s.l.) and shallow (<3 Km b.s.l.) plumbing system of Mt. Etna volcano.

## 2. Experimental strategy to simulate $CO_2$ flushing

The experimental strategy applied in this study was chosen to simulate the reaction of a partially crystallized $H_2O$-bearing magma with a fluid composed of $CO_2$ and $H_2O$ (with various $CO_2$ and $H_2O$ molar ratios). Two experimental steps were applied: (1) synthesis of a water-bearing (but water-undersaturated) partially crystallized material at 300 MPa and (2) reaction of this material with a fluid phase ~~to simulate fluid flux~~ at distinct pressure conditions of 300 and 100 MPa. The first step is aimed at reproducing the water-undersaturated basaltic magma ~~prevailing at depth~~ residing at high pressure into the storage system of Mt. Etna prior to $CO_2$ flushing. The second step simulates a fluid flux across the lower-intermediate plumbing system of Mt. Etna, where the presence of major magma storage zones at depths of 12-9 km and 6-3 km, together with a small reservoir at 1-2 km beneath the summit craters, have been inferred by geophysical studies (Aloisi et al. 2002; Patanè et al. 2006; 2013; Bonaccorso et al. 2011).

*2.1 Anhydrous glass synthesis*

A natural lava sample ET12_4M from the March 4, 2012 paroxysmal eruption at Mt. Etna was used as starting material (Table 1). This sample is a porphyritic K-trachybasalt ($SiO_2$ = 47.01 wt.%; Viccaro et al. 2015), which is among the most mafic materials collected at Mt. Etna during the post-2011 eruptive period (see Table 1 for major oxides composition of the sample; cf. Viccaro et al. 2015). This basalt contains 25 vol.% of phenocrysts in the following proportions: 40% clinopyroxene, 30% plagioclase, 15% olivine and 10% titaniferous magnetite. Phenocrysts are generally dispersed in a hyalopilitic groundmass.

The experimental starting material was prepared by crushing the natural lava to a grain size <5 mm. To further improve the homogeneity, an agate mortar was used to produce more fine-grained powder. The sample was melted for 3 hours in a Pt-crucible placed in a furnace at 1600°C and 1 atm. Then, the melt was quickly quenched to glass by quenching it directly on a brass plate. To improve the homogeneity of the mixture, the glass was grinded again in an agate mortar and melted for 2 hours at the same conditions. The major element composition and homogeneity of the final quenched basaltic anhydrous glass were examined by a Cameca SX 100 electron microprobe at the Institut of Mineralogy of the Leibniz Universität Hannover, Germany (Table 1). Analytical procedures for the Electron Microprobe Analysis (EMPA) are provided in Section 3.3. The glass was crushed and divided in two fractions with diameter of <200 μm and 200-500 μm. Before loading into experimental capsules, the two grain-size powders were mixed in a volume ratio of 1:1 to minimize the free volume between grains and hence, to reduce the incorporation of atmospheric water or nitrogen into the charge (Vetere et al. 2015a and references therein). This procedure ensures a more uniform distribution of fluids and avoids the incorporation of absorbed water in the sample container.

*2.2 First step experiment: sample preparation and starting material for the flushing experiments*

The glass powder prepared at 1 atm was first used to synthesize a hydrous glass containing 1 wt.% $H_2O$. A 40 mm long $Au_{80}Pd_{20}$-tube, with a diameter of 5.0-5.4 mm, was used to perform the experiments. ~~After annealing at 950°C, the capsule was prepared by loading~~ The capsule was annealed at 950°C and prepared for the synthesis of the hydrous glass by loading ≈1.0 g of glass powder and 1 wt.% of deionized water ~~for the synthesis of the hydrous glass~~. The capsule was ~~Capsules were~~ filled in several steps to achieve homogenous initial distribution of water and glass. Water was added step by step to dry glass powders and the charge was compressed after each addition of water by using a steel piston. Finally, the capsule was welded and checked for leakage of water by annealing at 110°C for about 2 hours. No water loss was detected by re-weighing the capsule. This capsule was used to synthesize a water-bearing glass at superliquidus conditions (1250°C, 300 MPa) in an internally heated pressure vessel (IHPV). The superliquidus conditions at 1250 °C at 300 MPa are confirmed using MELTS software and taking the average bulk rock composition of the March 2012 lavas of Mt. Etna as melt composition. After quench, the glass was analyzed for its water concentration by KFT (see below) and the remaining part was used again for step 1 experiment aimed at producing the starting material for $CO_2$-flushing experiments. The choice to hydrate the glass with only 1 wt.% of water relies on the intention to produce a starting material containing plagioclase at high pressure conditions. Such a choice also accounts for results of the hygrometer of Putirka (2008) applied on plagioclase cores from sample ET12_4M, which suggests an early crystallization of plagioclase at depth of 300 MPa, at ~1100°C with melt water content between 0.5 and 1.2 wt.%. The amount of added water also ensures that the melt is volatile-undersaturated (fluid-absent conditions) at the conditions of step 1 experiment (300 MPa, 1080°C).

To describe accurately textures obtained from the reaction of an assemblage (melt + mineral) in disequilibrium with fluids (simulating the fluid flux, see second step experiments), the approach of step 1 experiment was chosen to obtain particularly large crystals in the synthesis of the melt + mineral assemblage (step 1). The sample was prepared as described above for the synthesis of the water-bearing glass and brought to superliquidus conditions at 1250°C and 300 MPa for approximately one hour in an internally heated pressure vessel (IHPV). Subsequently, nucleation and crystallization processes were induced by lowering the temperature from liquidus conditions to 1080°C with a cooling rate of 20°C/min. Finally, 25 thermal cycles of two hours each with a temperature oscillating between 1060 and 1100 °C were applied. Pressure was kept constant at 300 MPa by automatic compensation. Before quenching, the temperature was set to the mean temperature (1080°C) for 20 minutes.

Crystallization under oscillating temperatures was done because this procedure decreases the number of crystals and enhances the size of crystals if the selected temperature interval is below the liquidus (cf. Hintzmann and Muller-Vogt 1969; Mills et al. 2011; Mills and Glazner 2013; Erdmann and Koepke 2016). When temperature is increasing (1060 to 1100 °C) crystals initially present at low temperature are partially dissolving and only the large crystals are expected to remain in the charge. When temperature is decreasing (1100 to 1060 °C) the crystallization is expected to occur mainly in the interface between the pre-existing crystals and the melt. The product obtained after the synthesis shows ~~indeed~~ crystals of clinopyroxene and plagioclase that are larger than 1 mm (Fig. 1). This product, labeled S0 in the following study, was used as starting material for the subsequent experiments in which a $H_2O+CO_2$ fluid mixture was added.

*2.3 Second step experiments: $CO_2$ flushing*

Two set of experiments were performed at two different pressures of 300 MPa and 100 MPa in order to reproduce $CO_2$-flushing at different levels of the Etnean plumbing system. Slices (14-20 mg) of the starting materials S0 were encapsulated in $Au_{80}Pd_{20}$-tubes together with deionized water and silver oxalate ($Ag_2C_2O_4$), which was used as source of $CO_2$. The ratio between the solid material S0 and the fluid phase ($H_2O+CO_2$) is given in Table 2. Water and silver oxalate were loaded into the capsules with different proportions to establish different mole fractions of $CO_2$ in the fluid composition at the onset of step 2 runs ($XCO_2^{fl}$ from 0.24 to 0.94; Table 2). Capsules were weighted after loading of each component and the welding; they were finally checked for leakage by annealing in oven for at least 1 hours at 110°C. For each set of experiments, three capsules were run simultaneously in order to evaluate the post-run compositional and textural changes for distinct fluid composition at exactly the same pressure and temperature conditions. The duration of each run was about 24 hours.

The P-T conditions selected for the fluid-bearing experiments (1080°C, 300 and 100 MPa) are consistent with those of the lower-intermediate plumbing system of Mt. Etna (Aloisi et al. 2002; Patanè et al. 2006; 2013; Bonaccorso et al. 2011). However, it must be noted that the experiments were performed at 1080°C, a temperature slightly lower than those generally assumed for Etnean K-trachybasaltic magmas at intratelluric condition (cf. Tanguy and Clocchiatti 1984). This slightly lower temperature enhances the crystallization of all mineral phases observed in the natural samples and leads to an increase of the crystal mean size.

## 3. Experimental and analytical methods

### 3.1 Internally heated pressure vessel (IHPV)

The IHPV apparatus used in this study is described in detail by Berndt et al. (2002). The temperature was measured by four unsheathed S-type thermocouples and an Eurotherm software was used to control temperature cycling and to record the measured temperature along the whole experimental run time. Precision of temperature of each thermocouple was ±10°C. Argon is the pressurizing medium and pressure was monitored by digital pressure transducers with an uncertainty of about 1 MPa. The variation of pressure during the experiments was <5 MPa. Experiments were performed at the intrinsic hydrogen fugacity of the vessel. Under intrinsic conditions in the IHPV, the oxygen fugacity in capsules containing $H_2O$-saturated melts (pure $H_2O$ fluid) was found to be close to the $MnO$-$Mn_3O_4$ or NNO + 3.7 buffer (Berndt et al. 2002). It must be taken into to account that, with increasing amount of loaded $CO_2$, the oxygen fugacity decreases due to the lowering of water activity, i.e., an increase in $XCO_2^{fl}$ from 0.1 to 0.9 decreases the $fO_2$ by roughly two logarithmic units (e.g., Botcharnikov et al. 2005, Vetere et al. 2014). The redox conditions prevailing within capsules were also determined on post-run products using the oxygen barometer of Ishibashi (2013), which is based on the spinel-melt equilibria (see analyses of glass and spinel in the electronic supplementary material ESM). The calculated log$fO_2$ varies between $10^{-6}$ to $10^{-7.3}$ bars, corresponding to a $fO_2$ ranging between NNO +2.9 and NNO +1.7 log units (Table 2). At the end of each run, samples were quickly quenched isobarically by dropping the capsule down into the cold part of the vessel, in order to avoid any non-equilibrium quench effects (rapid quench technique; estimated cooling rate was about 150°C/s; Berndt et al. 2002).

*3.2 Measurements of $H_2O$ content in initial hydrous glasses*

Karl-Fischer titration (KFT) method was adopted for the determination of the water concentrations in quenched initial glasses that were synthesized at superliquidus conditions and used for the first step experiment. Water is extracted by heating the sample up to temperature of about 1300°C and the released water is measured by coulometric titration. The coulometer used in this study is a Mitsubishi CA100. The benefit of the KFT method is that only small amounts of glass are required to get reliable analyses. In this study, about 50-100 mg of glass were required for water content of 1 wt.%. The extraction of water from the glass or melt with this technique can be incomplete in silica-rich glasses (Behrens 1995). Typically, about 0.10 wt.% of non-extracted $H_2O$ was found in samples containing initially more than 1.5 wt.% $H_2O$ (Behrens 1995; Leschik et al. 2004), implying that the water contents obtained with KFT methods may have to be corrected for values of residual water content. As the water diffusivity is much faster in silica-poor melts than in silica-rich melts (Behrens 1995; Behrens et al. 2004), the extraction of water should be greatly enhanced in the basaltic melts under investigation. Hence, the 1.1 wt.% $H_2O$ measured with KFT for ET12_M4 glasses can be assumed as representative of the real water content of the sample, and corrections for non-extracted water were not required. Considering that, after the cycling experiments the amount of crystallized phases [plagioclase, clinopyroxene and Fe-Ti oxide are 12.0, 22.9 and 3.8 area % respectively, calculated via image analyses as described in Vetere et al. (2015b)], mass balance calculations allow to infer the water content in the residual melt to a value of ≈1.5 wt.%. To test the accuracy of this value and provide a more accurate estimate of the initial water content for the equilibrium run S0, we used the plagioclase-liquid hygrometer of Waters and Lange (2015), which is based on the crystal-liquid exchange reaction between the albite and anorthite component and is applicable to a wide range of magmatic composition from basalts to rhyolites and alkaline magmas. Application of the hygrometer, using the temperature of 1080°C with the plagioclase (An58) and liquid composition of S0, results in a dissolved water content of 1.6± 0.35 wt.%, a value close to the previous estimate.

*3.3 Determination of sample fluid compositions after experiments*

The mole fractions of $H_2O$ and $CO_2$ in the fluid phase after experiments were calculated through mass balance using the initial amounts of loaded volatiles, rock powders and the estimated concentrations of volatiles retained in the melt after each experimental run. The $H_2O$ and $CO_2$ concentrations in the melt and the related composition of the fluid were determined using the MagmaSat software based on the thermodynamic model of Ghiorso and Gualda (2015), which allows us to calculate the saturation conditions for mixed $H_2O$-$CO_2$ fluids in natural silicate melt. In all the samples, the ~~$H_2O$-$CO_2$ degassing behavior~~ $H_2O$ and $CO_2$ partitioning between melt and fluid was modeled starting from the average composition of the run product glasses. The modelled $H_2O$ and $CO_2$ contents in the melts are in the range 3.06-0.73 wt.% $H_2O$ and 0.25-0.20 wt.% $CO_2$ for experiments performed at 300 MPa and 2.53-0.54 wt.% $H_2O$ and 0.07-0.03 wt.% $CO_2$ at 100 MPa (Table 2). Such concentrations are in the same range than those determined from melt inclusion analyses from other recent eruptions of Mt. Etna (Métrich et al. 2004; Spilliaert at al. 2006a; Collins et al. 2009).

*3.4 Electron Microprobe Analysis (EMPA)*

Major element compositions of the starting dry glass and the experimental products, composed of minerals and glass, were analyzed by a Cameca SX-100 electron microprobe at the Institute of Mineralogy of the Leibniz Universität Hannover, using 15 kV as the acceleration voltage. Conditions of measurement during glass analyses were 4 nA beam current and 20 μm diameter of electron beam. Following the procedure used at Hannover for the analysis of experimental charges containing hydrous basaltic glasses and minerals (e.g., Husen et al., 2016), peak counting times for glasses vary from 4 seconds for Na and K, to 8 seconds for the other elements. The analyses on minerals were conducted with a 15 nA beam current and a focused electron beam. Peak counting time was 5 seconds for Na and K, and 10 seconds for the other elements. Standard glasses and crystals were used for calibration. The composition of the starting dry glass is an average of 6 analyses. The average dry glass composition and related standard deviation is given in Table 1. Multiple point measurements were made for mineral phases and glasses of the experimental products to check the sample homogeneity and to reduce possible analytical errors. For some of the largest plagioclase crystals (~1 mm size), core-to-rim profiles were measured through single grains to determine changes in the chemical composition. Microprobe data for post-run minerals and glasses are synthetized in Table 3. The whole dataset is available as Electronic Supplementary Material (ESM hereafter).

## 4. Experimental Results

*4.1 Mineral assemblage and composition of the hydrated starting material ($S_0$)*

Hydrated products resulting from crystallization at 300 MPa (sample S0 in Figure1) consist of glass (~61 vol.%) and a mineral assemblage constituted by plagioclase (~12 vol.%) and clinopyroxene (~23 vol.%). Abundant and relatively small (20-150 μm in size) Fe-Ti oxides, mainly titaniferous magnetite, are present in proportion of ~4 vol.% of the sample (Fig. 1). The glass compositions, reported in Table 3, fall in the field of the basaltic trachyandesite ($SiO_2$ = 52-53 wt.%; $Na_2O+K_2O$ = 8.2-8.6 wt.%). As a whole, the observed mineral assemblage resembles that of the natural lava samples from Mt. Etna 2012 eruptions, with the exception of the lack of olivine crystals. The absence of olivine in S0 may be related to the rather oxidizing experimental conditions and to the effect of oxygen fugacity on the stability field of olivine. An increase of oxygen fugacity reduces the stability field of olivine (e.g., Roeder and Emslie 1970; Carmichael and Ghiorso 1990; Berndt et al. 2005; Feig et al., 2010).. Clinopyroxene is the most abundant mineral phase with size from 200 μm to about ~1 mm and with a diopsidic composition ($Wo_{47-50}En_{30-38}Fs_{14-21}$). The Mg# [atomic ratio Mg/(Fe+Mg)] varies between 77 and 88 (Fig. 2; electronic supplementary material, ESM hereafter). Clinopyroxene crystals display hollow cores and euhedral rims that indicate skeletal crystal growth possibly induced by the rapid initial cooling of the step 1 run. Plagioclase occurs as euhedral crystals up to 1 mm in size and is labradoritic in composition ($An_{55-64}$; Fig. 2; ESM). The chemical zoning of plagioclase is characterized by oscillations of the anorthite (An) content in a range of ~10 mol%. This feature is consistent with crystallization of plagioclase in a system with temperatures oscillating in a range of 40°C (Singer et al. 1995; Couch et al. 2001). It is worth noting that the average An content of plagioclase in the experimental sample is significantly lower than that of natural products of Mt. Etna, which typically covers a wider range from bytownite to labradorite, with subordinate, rare anorthite crystals (Viccaro et al. 2010; 2015; 2016). Textural relations between minerals show that most of the plagioclase nucleated when clinopyroxene and oxides were already formed, indicating that the growth of clinopyroxene and Fe-Ti oxides was faster than that of plagioclase (cf. Hammer and Rutherford 2002; Fig. 1).

*4.2 Mineral assemblage and mineral composition after $CO_2$-$H_2O$ flushing*

Textural characteristics of the sample S0 after fluid flushing at 1080°C are shown in Figure 3 and Figure 4 for experiments performed at 300 MPa and 100 MPa, respectively. At each pressure three experiments with different initial $XCO_2$ of the fluid phase were performed. Since the liquidus temperature of tectosilicates is strongly depended on the water activity (i.e., the liquidus temperature decreases with increasing water activity), plagioclase is found only in sample where $XCO_2^{fl}$ is high, as expected (Fig. 3e-f and Fig. 4c-f).

The high pressure experiments at 300 MPa produced a mineral assemblage with clinopyroxene and titaniferous magnetite for $XCO_2^{fl} = 0.4$ (sample 300_S2; Fig. 3a-b) and $XCO_2^{fl} = 0.7$ (sample 300_S3; Fig. 3c-d). At 300MPa and $XCO_2^{fl} = 0.4$ we observed relicts of clinopyroxene clearly affected by pervasive dissolution, which have compositions $Wo_{47-50}En_{33-38}Fs_{15-20}$ and average Mg# 85 (see Table 3; ESM). The crystal size is in the order of 300 to 400 μm (Fig. 3a-b) and is smaller than in the starting material S0. The composition of clinopyroxene is more evolved at $XCO_2^{fl} = 0.7$ ($Wo_{45-50}En_{32-37}Fs_{15-20}$; average Mg#81; Table 3; ESM), moreover, the maximum size is ~100 μm and crystals tend to have a more euhedral habitus than those observed at lower $XCO_2$ Fig. 3c-d). Oxides are 5-60 μm in size and constitute ~5% of the volume of the phenocryst phases; their compositions are comparable in both, At 300 MPa and $XCO_2^{fl} = 0.9$ (sample 300_S4) clinopyroxene and Fe-Ti oxides coexist with plagioclase (Fig. 3e-f). Plagioclase forms sub-euhedral crystals up to ~500 μm in size, commonly characterized by dissolution textures along their edges (compare Fig. 1 and Fig. 3f). Plagioclase composition is labradoritic ($An_{57-65}$) and rather similar to that in S0 ( Table 3; ESM; Fig. 2). Clinopyroxene is euhedral to sub-euhedral with composition $Wo_{46-49}En_{30-39}Fs_{14-22}$ and a size of ~300 μm. In contrast to plagioclase showing dissolution textures (Fig. 3f), clinopyroxene has rims usually marked by more Ca and Mg-rich envelopes (Mg# 87-89) enclosing cores. This composition more Mg-rich than that in S0 (Mg# 76-82; see ESM; Fig. 3e-f). The crystallinity of the products at 300 MPa increases progressively from ~10 vol.% in sample 300_S2 (high $H_2O$-low $CO_2$) to ~30 vol.% in 300_S4 (low $H_2O$-high $CO_2$). The higher crystallinity and lower crystal size of 300_S4 relative to S0 results in an increase in crystal number density. This feature can be explained by the lower melt water content in 300_S4 (0.74 wt.% $H_2O$) when compared with 300_S2 (1.60 wt.% $H_2O$) (Bugger and Hummer 2010; Riker et al. 2015).

At 100 MPa and $XCO_2^{fl} = 0.2$ (i.e. sample 100_S1), the product is only composed of glass ($SiO_2$ = 48 wt.%) and zoned clinopyroxene ($Wo_{47-50}En_{34-39}Fs_{13-17}$; average Mg# 0.91; see ESM; Fig.

2; Fig. 4a-b). The mineral assemblage changes to clinopyroxene, plagioclase and titaniferous magnetite at higher proportion of $CO_2$ in the fluid (i.e. 100_S2 and 100_S4; Fig. 4). For $XCO_2^{fl}$ = 0.2 clinopyroxene is euhedral to subhedral with some evidences of dissolution at the core. The shape of the clinopyroxenes is similar in the experiments at $XCO_2^{fl}$ = 0.2 and $XCO_2^{fl}$ = 0.5 (i.e. 100_S1 and 100_S2). However, the amount of clinopyroxene is clearly lower than in S0, indicating that dissolution and recrystallization probably occurred in these two experiments. At the highest $XCO_2$ (sample 100_S4), clinopyroxene shows skeletal textures similar to those of S0, though crystals greatly differ in size that is much smaller for the 100_S4 crystals (~200 μm; Fig. 4e-f); textural observations do not help to clarify if this mineral grew or dissolved. Despite the variations in crystal textures, the abundance and composition of clinopyroxene correlate with the $CO_2$ concentrations in the fluid. As the $XCO_2^{fl}$ of the system increases, the proportion of clinopyroxene becomes higher, whereas the average Mg# decreases from 0.91 (100_S1) to 0.84 (100_S4; Fig. 2; see ESM). When present, plagioclase occurs as 200 to 300 μm-large oscillatory-zoned crystals with composition ranging from $An_{53}$ to $An_{65}$ (Table 3; ESM). Observations under optical microscope and BSE images indicate that the size of plagioclases is much lower compared with the starting material S0 and, on average, it decreases by a factor of 2 in both the 300 MPa and 100 MPa experiments. Most plagioclase crystals have elongated euhedral habitus, but there are also crystals with subhedral to anhedral habitus, characteristic for dissolution textures (Fig. 4). Thus, textural observations indicate that plagioclase dissolved in all experiments at 100 MPa.

*4.3 Composition of the experimental glasses*

Major element composition of glasses record progressive melt evolution as a response of increasing $CO_2$ content in both sets of experiments (see ESM; Fig. 5). At a given pressure, $K_2O$ and $Na_2O$ become more concentrated in the melt with increasing $XCO_2$, while concentrations of MgO, CaO and $TiO_2$ decrease (Fig. 5). The comparison between the glass composition of sample S0 and that of the samples flushed with $CO_2$ is helpful to test the textural observations indicating dissolution or growth of the solid phases. Since $K_2O$ behaves almost as an incompatible element, $K_2O$ concentration in the residual glasses is a good proxy for tracing the crystallinity. Except for the experiment at 100MPa and 0.9 $XCO_2^{fl}$ (sample 100_S4), all charges record a significant decrease of $K_2O$ with respect to sample S0, indicating a decreasing bulk crystal content with increasing $XCO_2$ (Fig.5). For sample 100_S4, the $K_2O$ glass concentration is very close to that of S0, a feature that reflects a crystal/liquid proportion which is similar to that of the starting material.

A detailed analysis of the compositional variation of the elements other than $K_2O$ provides additional information on crystal growth or crystal dissolution during flushing experiments (Fig. 5). The $Na_2O$, $Al_2O_3$, $CaO$ and $MgO$ composition of glasses in samples 300_S2, and 100_S1 compared to S0 indicate that both clinopyroxene and plagioclase dissolved in these experiments at low $XCO_2$. The experiment 300_S3 (300 MPa, 0.7 $XCO_2^{fl}$) contains only clinopyroxene and no plagioclase, but the $Na_2O$ content of the glass in this experiment is slightly higher than that of the S0 (Fig. 5). This behavior can only be explained if a Na-poor phase (i.e., clinopyroxene) is crystallizing, as confirmed by the textural observation. The glass composition of sample 300_S4 (0.9 $XCO_2^{fl}$), which has lower MgO concentrations and higher $Na_2O$ and $Al_2O_3$ than S0, also implies that clinopyroxene is crystallizing and plagioclase is dissolving in this experiment. The glass composition of the experiment 100_S2 (0.5 $XCO_2^{fl}$) can only be explained if both phases are dissolving. In this case, evidence for plagioclase dissolution is given by textural observations, but the higher MgO content of the glass also indicates that clinopyroxene is slightly dissolving (Fig. 5). For the sample 100_S4 (0.9 $XCO_2^{fl}$), $K_2O$ indicates that the crystallinity is similar to S0. However, the slightly higher $Na_2O$ content of the glass in 100_S4 (at similar CaO, MgO and $Al_2O_3$ concentrations; Fig. 5) indicates that a small proportion of clinopyroxene is crystallizing and that a small proportion of plagioclase is dissolving. The evolution of most Fe-Ti oxides can be reconciled with crystallization or dissolution processes of the two minerals, i.e. clinopyroxene and plagioclase. However, FeO has a different behavior. The concentration in all glasses is in the same range (7.0 to 7.5 wt.% FeO) and is higher than that of S0 (Fig. 5). This higher FeO content is explained by the lower proportion of Fe-Ti oxides (dissolution) in the step 2 experiments, a feature also noted from textural observations. The dissolution of oxides is likely due to a slightly different oxygen fugacity in the S0 and in the step 2 runs (see Table 2).

## 5. Discussion

*5.1 Textural and compositional observations on minerals*

The comparison of the starting material S0 with the experiments performed at 300 and 100 MPa at various $XCO_2^{fl}$ gives us a direct link to understand the role of $CO_2$ on the phase stability of basaltic systems enriched in carbon dioxide, such as Mt. Etna. In particular, we observed that the high $a_{H2O}$ in the system during crystallization of Etnean basalts at 300 MPa caused complete dissolution of plagioclase in samples 300_S2 (3.06 wt.% $H_2O$ in the melt) and 300_S3 (1.90 wt.% $H_2O$ in the melt). Conversely, flushing the system with fluids with high $CO_2$ proportions at high pressure (run 300_S4, 0.73 wt.% $H_2O$ in the melt), strongly destabilizes plagioclase, which develops dissolution textures despite the water content of the system is significantly reduced with respect to S0 (Fig. 1 and Fig. 3e-f). At 100 MPa plagioclase crystals are present in samples 100_S2 and 100_S4 for melt water contents between 1.78 and 0.54wt. %. These crystals have subhedral habitus with minor evidence of dissolution surfaces in both 100_S2 and 100_S4 samples (Fig. 4c-f).. Even more interesting is the fact that the amount of water in melts of the plagioclase-bearing charges 300_S4 and 100_S4is significantly lower than in the starting material S0 (see Table 2), a condition that should promote the plagioclase stability and enhance crystal growth according to a number of experimental studies (e.g., Housh and Luhr 1991; Lange et al. 2009; Almeev et al. 2012). Nevertheless we observed that plagioclase is dissolving in these two experiments.

Different proportions of $H_2O$ and $CO_2$ in the experimental charges also produce major variations of textures of clinopyroxene that are ubiquitous in all the samples. Textures of clinopyroxene crystals vary at increasing $XCO_2^{fl}$, with different behavior depending on pressure. In the high pressure experiments the clinopyroxene is destabilized for high melt water concentration (3.0 wt.% $H_2O$, sample 300_S2; see Table 2), but its stability gradually expands at increasing $CO_2$ in the fluid phase; under these conditions clinopyroxene crystals develop Ca- and Mg-rich growth zones towards the rim (Fig. 3e-f). Low-pressure charges at high water concentration in the melt and low $XCO_2^{fl}$ display euhedral grains of clinopyroxene (sample 100_S1, 2.5 wt.% $H_2O$), whereas crystals with hollow cores and anhedral rims are characteristic for charges flushed with a fluid with high $CO_2$ proportions ( sample 100_S4, 0.54 wt.% $H_2O$). .However, comparison between the starting material S0 and experimental samples of run 2 indicate increasing modal proportion of clinopyroxene for elevated proportion of $CO_2$ during both the high and low pressure experiments.

Despite the evidence of textural changes for plagioclase and clinopyroxene, the incorporation of large amount of $CO_2$ in the starting material does not produce significant compositional changes on the entire phase assemblage that preserve compositions very close to that of S0 (Fig. 2). As a whole, we observed that the effect of changing $CO_2$ proportions in basaltic melts is small on clinopyroxene Mg# at 300 MPa (ΔMg# ~2%). Clinopyroxene records more prominent chemical changes at 100 MPa; the Mg# correlates negatively with $XCO_2$, shifting from 0.84 for $XCO_2^{fl}$=0.9 to 0.91 for $XCO_2^{fl}$=0.2. Plagioclase behaves similarly at 100 MPa, with the average An content of plagioclase increasing slightly from $An_{58}$ at $XCO_2^{fl}$=0.9 to $An_{60}$ at $XCO_2^{fl}$=0.5. However, the chemical variation of plagioclase may not be representative, given that analytical errors are on the order of 2 mol% An.

On the whole, the mineral chemistry of experimental products is similar to products emitted at Mt. Etna both in historic and recent eruptions. With the exception of the clinopyroxene crystals obtained very low $XCO_2$, the composition of the experimental clinopyroxene is in the range of compositions of the natural samples, which falls in the fields of diopside-salite and augite, and has average Mg# between 0.79 and 0.83 (Giuffrida and Viccaro, 2017). Similar observations can be done for plagioclase: the experimental phases fit the compositional range observed for plagioclase crystals found in Etnean natural samples [$An_{47-91}$, with frequency peaks at the labradoritic compositions for crystals of the historic and recent activity (Viccaro et al., 2010; 2016)].

*5.2 Reaction of the mineralogical assemblage on $CO_2$ flushing*

The isobaric and isothermal experimental series presented in this paper simulate the reaction of a magma as a result of changing $H_2O$ and $CO_2$ proportions in the volatile mixture only (constant temperature and pressure). The classical phase equilibria studies in mafic systems in which one bulk composition is in equilibrium with a $H_2O$-$CO_2$–bearing fluid phase all indicate that the phase proportions of both plagioclase and clinopyroxene should decrease with increasing water activity at fixed temperature and pressure (e.g., Feig et al. 2006; Pichavant and Mac Donald 2007; Almeev et al. 2013). The analysis of previous studies shows that the same general observation can be done in pure $H_2O$-bearing systems (no $CO_2$): reduction of the water concentration in the melt promotes the plagioclase and clinopyroxene stability and enhances crystallization (e.g., Moore and Carmichael 1978; Blundy and Cashman 2001; Berndt et al. 2005; Almeev et al. 2012). Our experiments indicate (1) that the effect of $H_2O$ on the stability and mineral proportion of plagioclase is much more pronounced than that of $H_2O$ on clinopyroxene, a result that is in agreement with past experimental studies and (2) that clinopyroxene is crystallizing and plagioclase is dissolving in some experiments. This unexpected last observation cannot be reconciled with the previous phase equilibria studies and needs a detailed discussion.

Crystallization of clinopyroxene and dissolution of plagioclase was observed in two experiments at 300 MPa and one experiment at 100 MPa. In two of these experiments (run products at very high $XCO_2$), the water content of the melt is clearly lower than the initial water content (estimated to be 1.6 wt% $H_2O$; Table 2). The third experiment at 300 MPa has approximately the same $H_2O$ content of the melt than the starting material prepared at the same pressure (~1.9 wt% $H_2O$). Compared with the general results obtained in pure $H_2O$-bearing systems, the crystallization of clinopyroxene in the systems with lower $H_2O$ content of the melt than the starting material can be expected. Thus, the behavior of clinopyroxene in the flushing experiments is consistent with previous experiments but the dissolution of plagioclase is problematic and difficult to explain. The experimental approach used in our two step experiments implies that a $CO_2$–free assemblage (first step) is reacting with a $H_2O$-$CO_2$–bearing fluid (second step). Thus, at high $XCO_2$, $H_2O$ is released from the melt and $CO_2$ is incorporated into the melt phase. Considering that the decrease of water content only is not expected to cause a dissolution of plagioclase (see above), another parameter must be taken into account, especially in the 300 MPa experiments because both steps (synthesis of the starting material and flushing experiment) were conducted at the same pressure. In addition to changing melt water content, three explanations can be proposed for the plagioclase dissolution observed in the 300 MPa experiments: disequilibrium in the S0 and/or in step 2 experiments, oxygen fugacity and the $CO_2$ content of melt.

*Disequilibrium due to crystallization kinetics in S0 run:* The shape of the minerals observed in the S0 experiments (Fig. 1) is not that expected for minerals in textural equilibrium with a melt phase. Thus, it may be possible that equilibrium was not attained in the S0 experiment. The shape and the large size of the minerals in the S0 run is related to the temperature path of the experiment, which started from an initial temperature of 1250°C at conditions above the liquidus. At this stage a homogeneous melt phase was formed and nucleation within a homogeneous melt is known to be difficult, in contrast to experiments starting directly from a glass powder. Thus, with a cooling rate of 20°C/min, the growth rate of the few minerals that nucleated is expected to be fast, probably leading to minerals with a squelettal shape. The subsequent temperature cycling explains the compositional variations of the large minerals. However, a detailed analysis of Figure 1 indicates that the evidence of a squelettal shape is observed in the internal part of the crystals, but the outer part of the mineral phases is in textural equilibrium (the typical crystalline orientations of plagioclase and clinopyroxene can be recognized). Considering the sluggish nature of plagioclase, it is possible that equilibrium proportions of plagioclase were not reached in the S0 experiments (~ 50 hours). However, in this case, unexpected crystallization of plagioclase would occur in the step 2 experiments, but not dissolution as observed in this study. Thus, problems related to the kinetics of plagioclase growth in S0 runs cannot explain the experimental observations.

*Diffusivity of volatiles in step 2 runs:* The step 2 experiments were conducted by equilibrating a fluid phase with a 14-20 mg piece of partially crystalline sample. Since $H_2O$ can diffuse faster than $CO_2$ in silicate melts (e.g., Zhang and Ni, 2010) the dissolution of plagioclase could be explained by disequilibrium between the volatile concentration in the melt and the fluid in the first stages of the experiment. In this case, the water activity in the central part of the samples may be higher than expected leading to dissolution of plagioclase, but dissolution textures should not be observed in rim of the samples. This contrasting behavior in the rim and the core of the samples was not observed. In addition, the chips used for the step 2 experiments are small and equilibrium is expected to be reached after 24 hours in depolymerized basaltic melts. However, considering the sluggish reaction of plagioclase, this explanation cannot be definitely excluded.

*Oxygen fugacity:* In IHPVs, the decrease in water activity is accompanied by a decrease in oxygen fugacity, as described by Scaillet et al. (1992) and Botcharnikov et al. (2005). However, phase equilibria in mafic systems at relatively low water contents do not indicate that the plagioclase stability is strongly affected by changing redox conditions in the $fO_2$ range explored in this study. At water activity close to unity, the liquidus temperature of plagioclase may be slightly depressed with decreasing $fO_2$ in trachytic or granodioritic systems (e.g. Dall'Agnol et al. 1999; Martel et al. 2013). Results obtained in basaltic and granodioritic systems at water activities below 0.5, which may be more relevant for comparison with this study, do not show a strong influence of oxygen fugacity on the beginning of crystallization of plagioclase (e.g. Dall'Agnol et al. 1999; Berndt et al. 2005; Botcharnikov et al. 2008) and of clinopyroxene (Botcharnikov et al. 2008). However, even if strong effects were not detected within 50 °C intervals (usually the temperature intervals for the phase equilibria were 50 °C) oxygen fugacity is expected to cause structural changes in the melt by changes of the $Fe^{2+}/Fe^{3+}$ ratio, which should at least affect slightly the stability of clinopyroxene, magnetite, and plagioclase.

Except for the results presented in this study, there is no experimental data allowing us to compare the stability of plagioclase at constant melt $H_2O$ content between $H_2O$-$CO_2$–bearing systems and $H_2O$-bearing systems. Since changes of phase stabilities due to variations of $fO_2$ only may not be sufficient to explain the experimental observations, our results indicate that the incorporation of small amount of $CO_2$ in silicate melts as carbonate species influences significantly plagioclase stability (e.g., preferential bonding of Ca with carbonates in the melt phase can affect plagioclase stability). This hypothesis is consistent with the differences observed at 100 and 300 MPa. Our experimental results indicate that dissolution of plagioclase is more pronounced at 300 MPa than at 100 MPa, which may be related to the higher concentration of dissolved $CO_2$-bearing species at high pressure (e.g., Shishkina et al. 2014b; Vetere et al. 2014). Considering the $H_2O$ and $CO_2$ solubility data in silicate melts from Etna (Shishkina et al. 2014b), approximately 3000 ± 500 ppm $CO_2$ should be incorporated in Etna basalts at 300 MPa and at high $XCO_2$. In our experimental melts, this concentration may reach 5000 ppm or more, considering that the melts obtained at high $XCO_2$ were more alkali-rich than the starting melts and that $CO_2$ solubility is increasing with increasing alkalis (Vetere et al. 2014). Since $Fe^{2+}/FeO_{tot}$ ratio is directly proportional to the amount of $CO_2$ dissolved in melts (Vetere et al. 2014), $fO_2$ changes to lower values (or more reduced conditions) may also increase the amount of dissolved $CO_2$. However, if our hypothesis is correct, the strong effect of such low $CO_2$ concentrations on the plagioclase stability is surprising and may be important to understand the effects of $CO_2$ fluxes in magmatic systems.

The most reliable explanation for the textural observations made in this study is a combination of structural effects resulting from $CO_2$ incorporation and from changes of activities of the individual components due the different effects of volatiles ($H_2O$ and oxygen fugacities) on the stability of solid phases (e.g. $H_2O$ has a stronger effect on the stability of plagioclase than on the stability of clinopyroxene). However, although not detected in the experimental products, kinetic effects due to different diffusivities of $H_2O$- and $CO_2$-bearing species in the basaltic melts cannot be fully excluded. However, additional systematic tests should be performed to confirm the results of our pilot experiments.

## 6. Implications and conclusions

Empirical investigations aimed at reproducing the interaction between silicate melts and volatile species, where the fluid is realistically multicomponent, i.e. $H_2O+CO_2$, are crucial for interpreting pre-eruptive history of magmas that are largely enriched in $CO_2$, such as those of Mt. Etna. Our experiments on Etnean K-trachybasalts directly simulate processes of $CO_2$-flushing at distinct sections of the plumbing system, providing new insights for interpreting the range of crystal textures observed in natural samples.

Although compositions of the experimental mineral phases do not significantly change with $CO_2$, we have evidence that plagioclase and clinopyroxene react differently as a consequence of the addition of significant amounts of $CO_2$. Under elevated $CO_2$ proportion with respect to water, plagioclase crystals may experience severe dissolution at 300 MPa, whereas the destabilization of plagioclase is less pronounced at 100 MPa. Clinopyroxene textures change differently during the high and low pressure experiments of flushing. Clinopyroxene stability expands and crystals display growth textures at 300 MPa and high $XCO_2$; in contrast, clinopyroxene growth is much less pronounced at low pressure for comparable volatile concentration.

Because $CO_2$ may have a significant influence on magma crystallinity, the effects of $CO_2$-flushing on products from the recent activity of Mt. Etna cannot be neglected for a proper interpretation of pre-eruptive dynamics. In Etna basalts, the presence of plagioclases characterized by cores with dissolution/resorption textures, such as corse-sieve and patchy textures, is common (Viccaro et al. 2010; 2014; 2015; 2016; Giuffrida and Viccaro 2017 Fig.7). Moreover, the rims of plagioclase are often affected by single or multiple sieve textures (Fig.7). Most of the augitic clinopyroxenes that were found in lava rocks from the post 2011 paroxysmal activity (e.g. eruptions of March 04, 2012, and April 18, 2013; cf. Viccaro et al. 2014, 2015; Giuffrida and Viccaro 2017) also display destabilization along the rim, sometimes evolving to pervasive disequilibrium features similar to sieve textures (Fig.7). The development of disequilibrium features at the crystal rims has been often attributed to processes of recharge of the residing system by more basic magma, whereas decompression of water-undersaturated magmas has been thought to be the main process controlling the destabilization of crystal cores (Viccaro et al. 2010, 2014; Giuffrida and Viccaro, 2017). Experimental results presented in this paper demonstrate that the reaction of crystal-bearing magmas with $CO_2$–rich fluids may also be responsible for the range of disequilibrium textures observed in the crystal cores and rims.

Exploring the degassing path of Etnean magmas variably enriched in $H_2O$ and $CO_2$, various authors concur on a $CO_2$ exsolution depth >400 MPa, i.e. a depth greater than the depth of the plagioclase crystallization threshold which is located at ~12 km, corresponding to ~340 MPa (Viccaro et al. 2010). Conversely, water is kept almost entirely dissolved in the melt up to ~5.5-7 km (~155-200 MPa). This means that, at the highest pressure investigated by our experiments, i.e. 300 MPa, the system is already volatile-saturated. Although the water concentration in the melt is probably enough to produce plagioclase dissolution/resorption during magma decompression below depths of 5.5–7 km, the migration of CO2-rich bubbles that are released from deeper regions of the plumbing system (> 400 MPa) should further promote plagioclase destabilization during their early growth history. Under these circumstances, the spectrum of disequilibrium textures at variable resorption degree that affect the coresof Etnean plagioclases may form in response of different concentrations of CO2 in the exsolved gas mixture flushing the magmatic system in which crystals grow.

On the same line of evidence, the transfer and injec- tion of prevalent CO2-rich gas into the shallow reservoirs (< 100 MPa) may have reduced the clinopyroxene stabil- ity, causing severe destabilization of their rims. Even in this case, the development of more pervasive disequilibrium features in clinopyroxenes, such as the sieve textures, is consistent with a possible increase of CO2 concentration in the gas phase flushing the magma stored at shallow crustal levels. Such a process may also explain the occurrence of some sieve textures in plagioclase crystals that cannot be associated to processes of mafic recharge due to the lack of correlation between anorthite, iron and/or magnesium con- tents (Viccaro et al. 2014).

**Acknowledgements**

MG acknowledges the PhD fellowship and two PhD research grants from the University of Catania. This work was also supported by a FIR 2014 project from UniCt to MV (grant number 2F119B). F. Vetere wishes to acknowledge the European Research Council (ERC) Consolidator Grant ERC-2013-CoG No. 612776 – CHRONOS and Marie Curie Fellowship IEF SolVoM #297880 to Diego Perugini.

**Figure Captions**

**Figure 1** – BSE image of sample S0 that has been used as starting material for experiments of flushing with water and carbon dioxide. S0 has been crystallized from a melt hydrated with 1 wt.% of water. Experiment performed with thermal cycling at 1080°C±20°C and pressure of 300 MPa. Cpx = clinopyroxene; plg = plagioclase; ox = opaque oxides.

**Figure 2** – Variation of the average Mg# of clinopyroxene and An (mol.%) of plagioclase crystals in the initial sample S0 and for the products flushed at 100 and 300MPa using a fluid with distinct proportion of $H_2O+CO_2$. $XCO_2^{fl}$ is the mole fraction of $CO_2$ in the fluid phase. Bars on data indicate the 2 sigma error.

**Figure 3** – Textural characteristic of sample S0 after flushing with a fluid composed of $H_2O+CO_2$ at 300MPa and 1080°C. Experimental results are shown for increasing proportion of $CO_2$ in the system ($XCO_2^{fl}$). Note that in condition of high water activity (lower $XCO_2^{fl}$) the mineral assemblage is only constituted by clinopyroxene (cpx) and opaque oxide (ox) in samples 300_S2 and 300_S3. Plagioclase (plg) is found in sample 300_S4, which was flushed with a gas phase at higher $CO_2$ concentration. Extensive dissolution features can be observed along the plagioclase rims.

**Figure 4** – Textural characteristic of sample $S_0$ after flushing with a fluid composed of $H_2O+CO_2$ at 100MPa and 1080°C. Experimental results are shown for increasing mole fraction of $CO_2$ in the fluid phase, as in Figure 3. Note that as the $XCO_2^{fl}$ increases, clinopyroxene crystals show major textural changes and higher degree of resorption, whereas plagioclase crystals do not record significant textural variation.

**Figure 5** – Major element compositions of residual glasses for increasing $CO_2$ composition after flushing experiments at 300MPa and 100MPa. In diagrams the composition of the glass in sample S0 is also shown for comparison.

Table 1. Bulk rock composition of the natural starting material ET12_4M (wt.% anhydrous) determined by Viccaro et al. (2015) and average major oxide composition (wt.%) of the experimental dry glasses and of the starting S0 glass (step 1 experiment) with standard deviations.

|  | $SiO_2$ | $TiO_2$ | $Al_2O_3$ | $FeO_{tot}$ | $MnO$ | $MgO$ | $CaO$ | $Na_2O$ | $K_2O$ |
|---|---|---|---|---|---|---|---|---|---|
| **ET12_4M** | 47.01 | 2.02 | 16.28 | 12.09 | 0.22 | 5.61 | 11.00 | 3.33 | 2.00 |
| **Dry glass*** | 47.37 ±0.42 | 1.83 ±0.03 | 16.55 ±0.34 | 9.86 ±0.34 | 0.13 ±0.03 | 5.42 ±0.10 | 10.60 ±0.23 | 3.34 ±0.69 | 1.90 ±0.04 |
| **S0 glass**** | 54.73±0.03 | 1.55±0.02 | 19.01±0.13 | 6.12±0.18 | 0.23±0.01 | 3.49±0.06 | 5.91±0.31 | 5.26±0.22 | 3.71±0.03 |

*average values calculated on 6 electron microprobe analyses

** average values calculated on 5 electron microprobe analyses

Table 2. Experimental conditions, composition of the fluid phase dissolved and exsolved after runs and post run products. Run $S_0$ at 1080±20°C ; all the other runs at 1080°C.

| Run | P (MPa) | log $fO_2$ | Solid fraction (mg) | Initial fluid fraction $XCO_2^{fl}{}_{In}$ | Initial fluid compositions (wt.%) | | Fluids in the melt (wt.%)* | | Exsolved fluids (wt.%)* | | Final fluid mole fraction $XCO_2^{fl}{}_{fi}$ | Post-run products** |
|---|---|---|---|---|---|---|---|---|---|---|---|---|
| | | | | | $H_2O$ | $CO_2$ | $H_2O$ | $CO_2$ | $H_2O$ | $CO_2$ | | |
| $S_0$ | 300 | -6.95 | 1000 | 0 | 1.1 | 0 | n.a. | n.a. | n.a. | n.a. | 0 | Cpx+Plg+Ox |
| 300_S2 | 300 | -6.62 | 20 | 0.43 | 3.9 | 7.3 | 2.63 | 0.22 | 1.27 | 7.08 | 0.69 | Cpx **diss** + Ox |
| 300_S3 | 300 | -6.79 | 15 | 0.74 | 3.5 | 24.9 | 1.72 | 0.26 | 1.78 | 24.64 | 0.85 | Cpx *cryst* + Ox |
| 300_S4 | 300 | -7.26 | 14 | 0.94 | 0.7 | 29.5 | 0.5 | 0.28 | 0.2 | 29.22 | 0.98 | Cpx **cryst** + Plg **diss** +Ox |
| 100_S1 | 100 | -7.01 | 17 | 0.24 | 4.9 | 3.8 | 2.39 | 0.03 | 2.51 | 3.77 | 0.38 | Cpx **diss** |
| 100_S2 | 100 | -6.82 | 14 | 0.51 | 3.3 | 8.3 | 1.58 | 0.05 | 1.72 | 8.25 | 0.66 | Cpx *diss* + Plg **diss** + Ox |
| 100_S4 | 100 | -6.04 | 15 | 0.94 | 0.7 | 28.8 | 0.39 | 0.09 | 0.31 | 28.7 | 0.97 | Cpx *cryst* +Plg *diss* +Ox |

*Composition calculated through MagmaSat software (Ghiorso and Gualda, 2015).

**In bold: strongly dissolving or crystallizing; in Italics: slight dissolving or crystallizing

Fig. 1

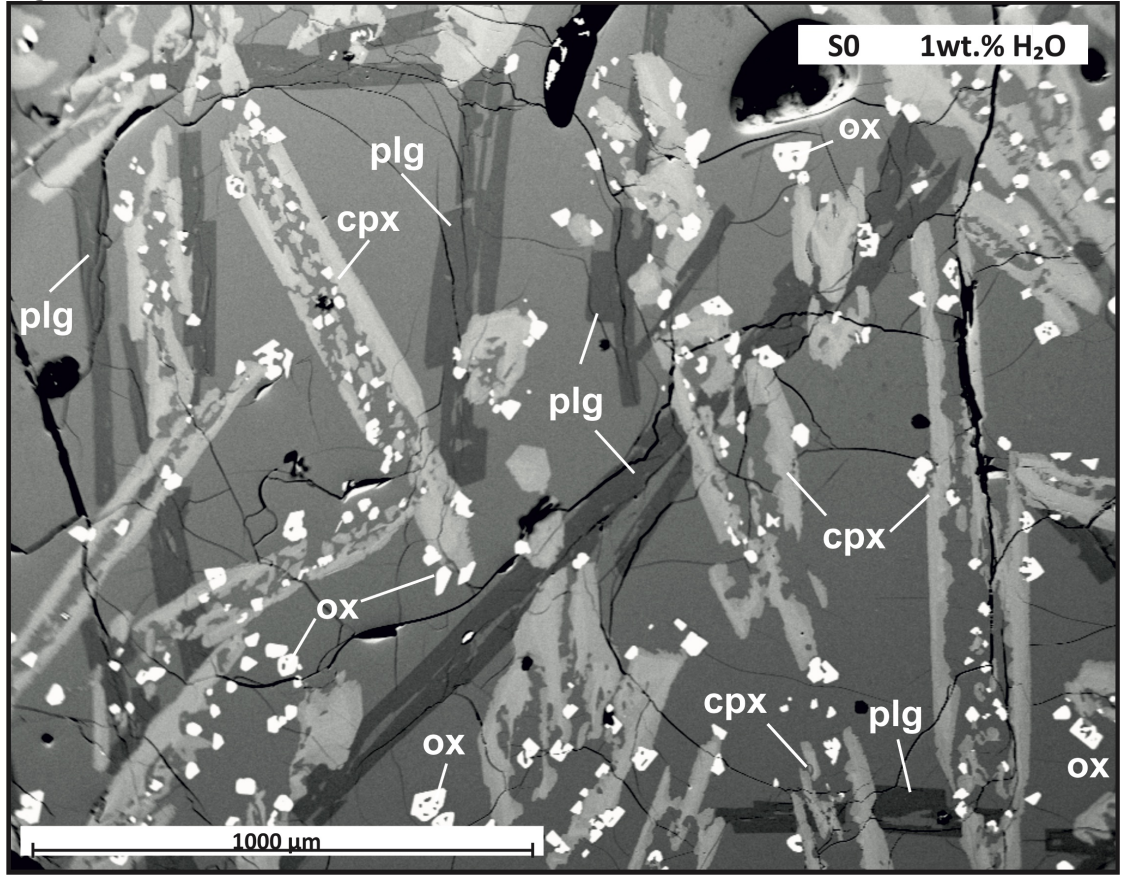

Fig. 2

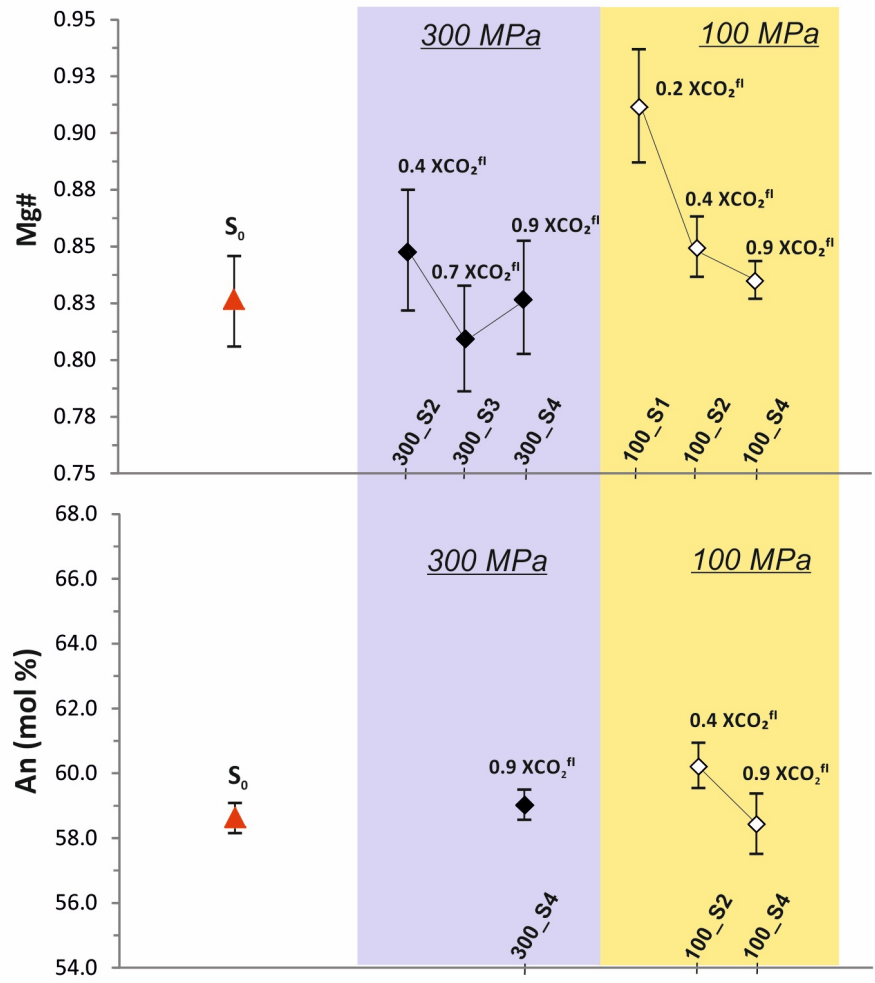

Fig. 3

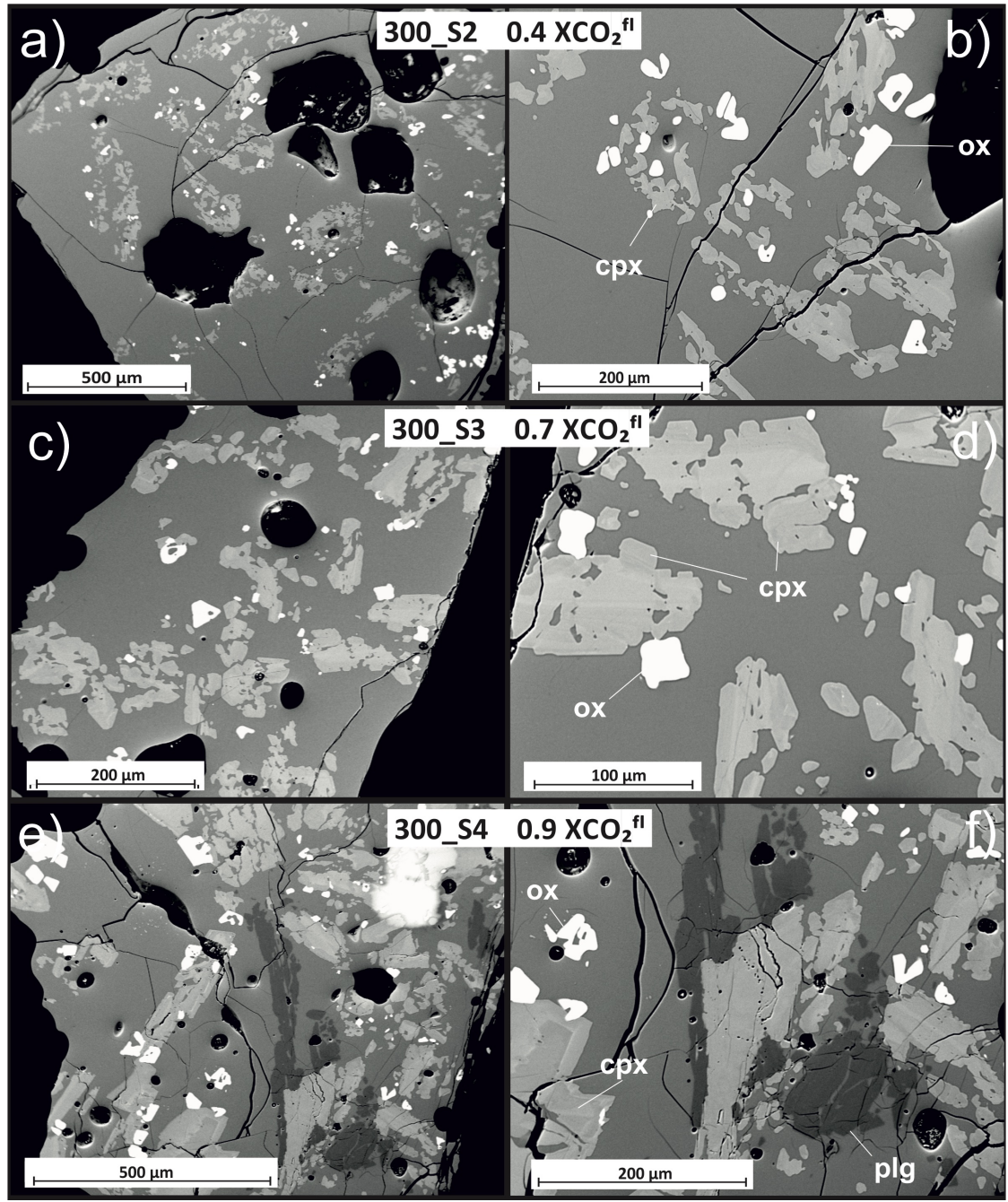

Fig. 4

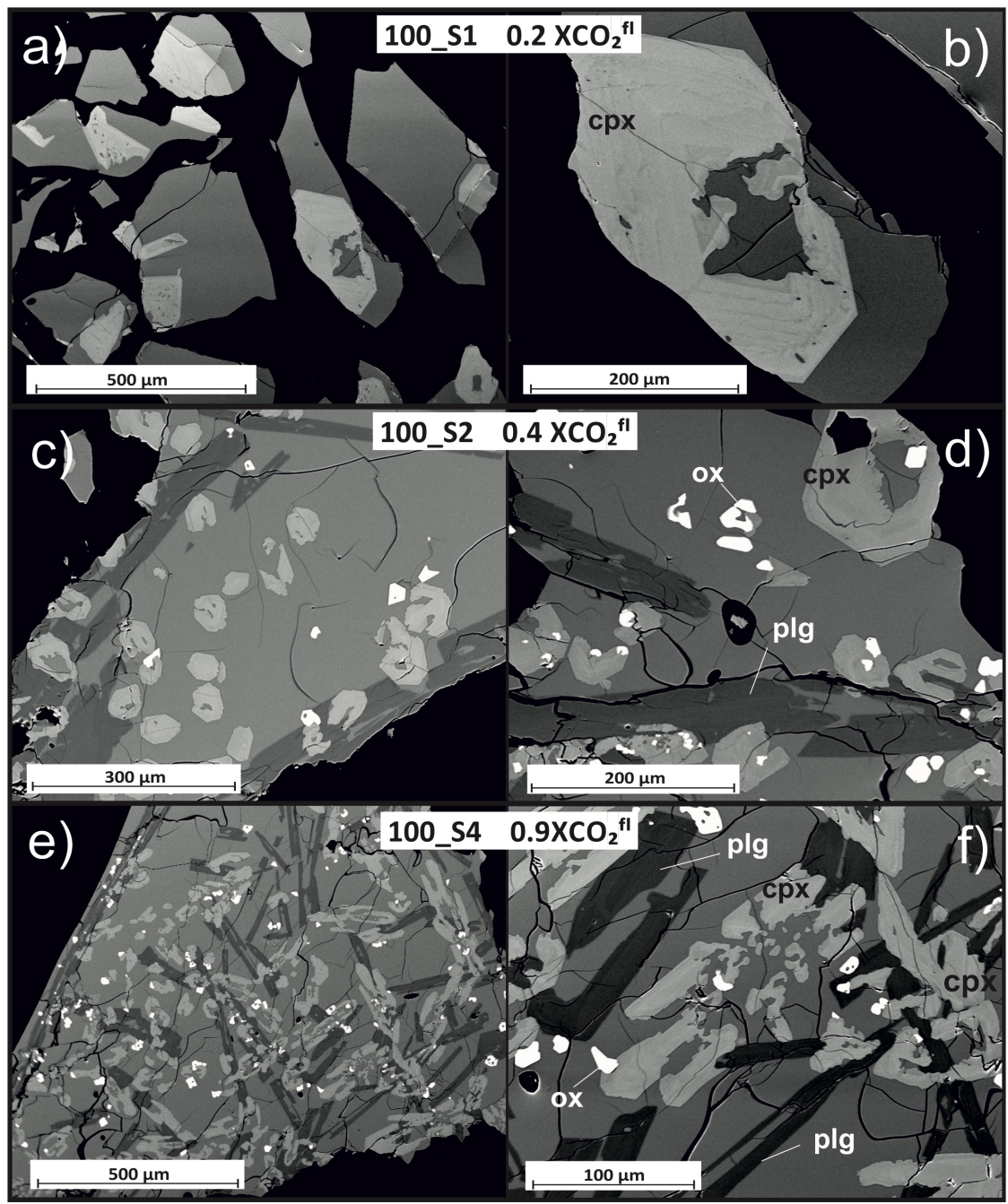

Fig. 5

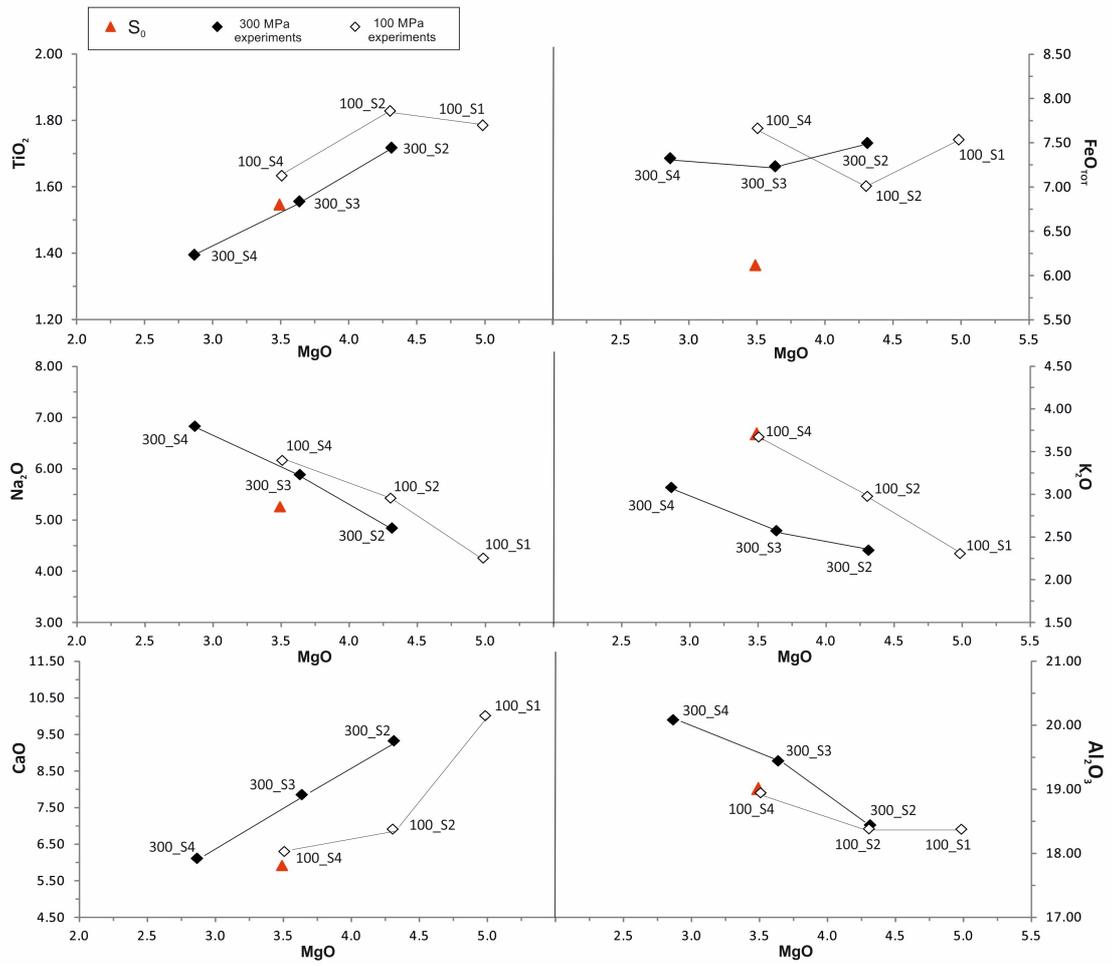

Fig. 6

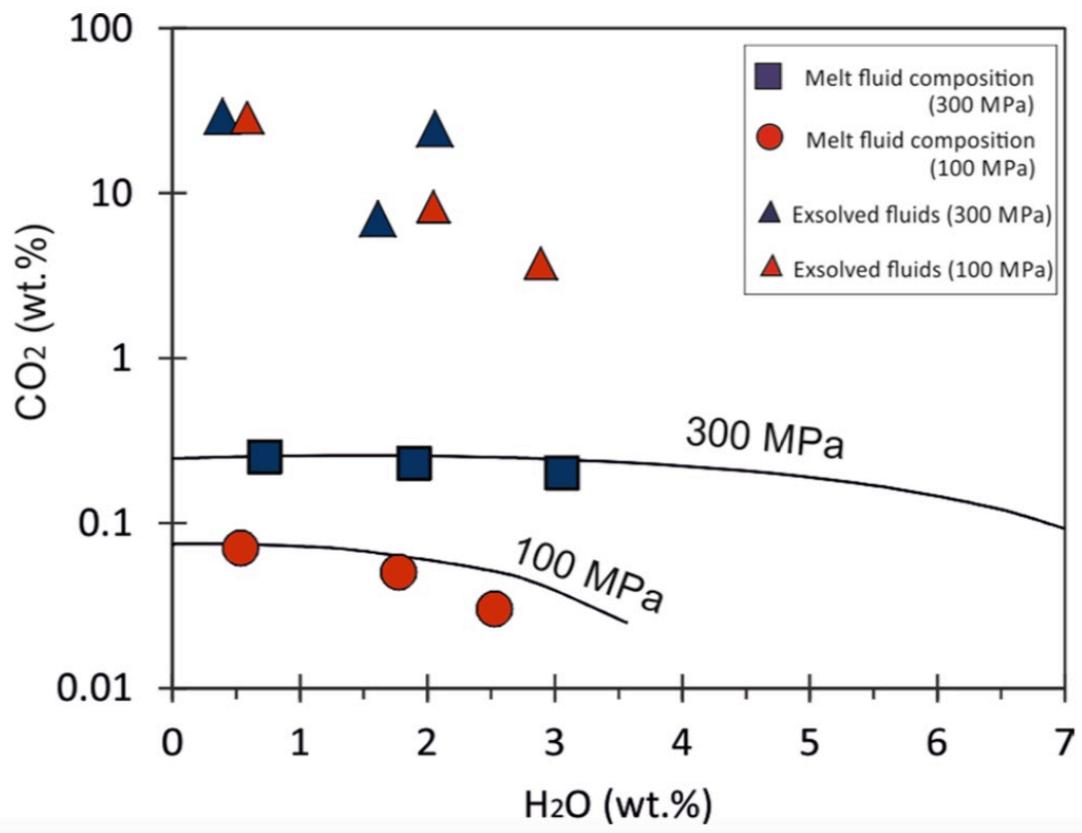

Fig. 7

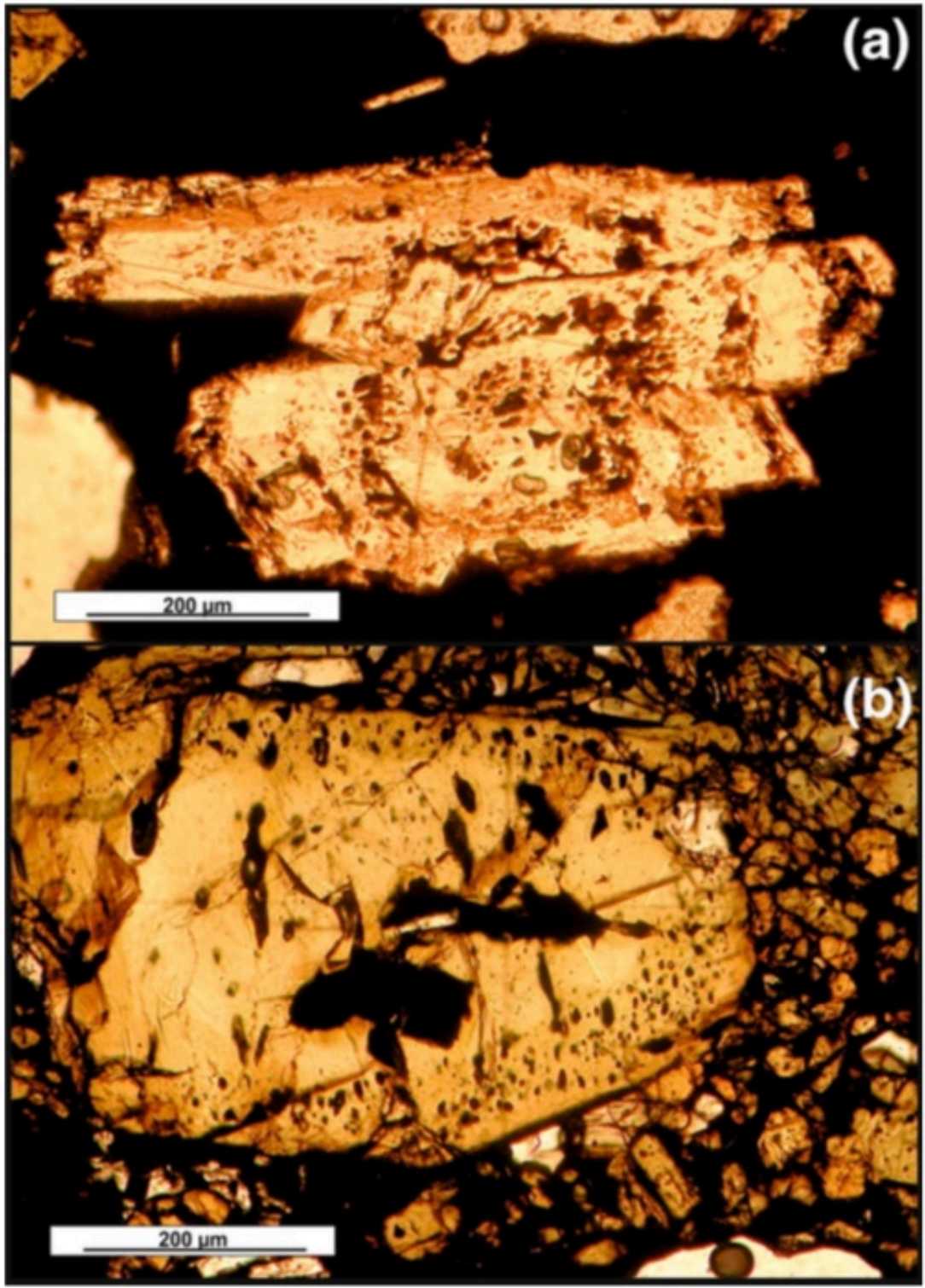

Major oxide composition of the experimental glasses

### H2O-bearing system
*S0*
Experimental conditions: 1 wt.% H2O; Thermal Cycling  76h  1080±20 °C ;  P=300 MPa.

| spots | SiO$_2$ | TiO$_2$ | Al$_2$O$_3$ | FeO | MnO | MgO | CaO | Na$_2$O | K$_2$O | Total |
|---|---|---|---|---|---|---|---|---|---|---|
| A1 | 54.71 | 1.56 | 19.19 | 5.96 | 0.24 | 3.53 | 5.67 | 5.36 | 3.78 | 100.00 |
| A2 | 54.61 | 1.46 | 18.93 | 6.14 | 0.24 | 3.62 | 6.32 | 5.04 | 3.64 | 100.00 |
| A3 | 54.74 | 1.63 | 19.18 | 5.88 | 0.20 | 3.57 | 5.96 | 5.03 | 3.80 | 100.00 |
| A4 | 54.84 | 1.52 | 18.98 | 6.19 | 0.29 | 3.31 | 5.98 | 5.26 | 3.63 | 100.00 |
| A5 | 54.75 | 1.56 | 18.77 | 6.42 | 0.16 | 3.44 | 5.63 | 5.59 | 3.69 | 100.00 |
| **Average** | **54.73** | **1.55** | **19.01** | **6.12** | **0.23** | **3.49** | **5.91** | **5.26** | **3.71** | **100.00** |

### H2O+CO2-bearing system
*300_S2*
Experimental conditions:  0.4 XCO$_2^{fl}$; P=300 Mpa; T 1100°C

| spots | SiO$_2$ | TiO$_2$ | Al$_2$O$_3$ | FeO | MnO | MgO | CaO | Na$_2$O | K$_2$O | Total |
|---|---|---|---|---|---|---|---|---|---|---|
| A2 | 51.42 | 1.73 | 18.49 | 7.29 | 0.02 | 4.31 | 9.59 | 4.84 | 2.32 | 100.00 |
| A3 | 51.33 | 1.86 | 18.71 | 7.49 | 0.14 | 4.33 | 9.29 | 4.33 | 2.52 | 100.00 |
| A4 | 51.55 | 1.64 | 18.67 | 7.66 | 0.12 | 4.27 | 9.04 | 4.79 | 2.25 | 100.00 |
| A5 | 51.97 | 1.64 | 17.95 | 7.59 | 0.04 | 4.08 | 9.38 | 4.91 | 2.43 | 100.00 |
| A6 | 50.80 | 1.72 | 18.36 | 7.45 | 0.24 | 4.58 | 9.33 | 5.33 | 2.19 | 100.00 |
| **average** | **51.42** | **1.72** | **18.44** | **7.49** | **0.11** | **4.31** | **9.32** | **4.84** | **2.35** | **100.00** |

*300_S3*
Experimental conditions:  0.7 XCO$_2^{fl}$; P=300 Mpa; T 1100°C

| spots | SiO$_2$ | TiO$_2$ | Al$_2$O$_3$ | FeO | MnO | MgO | CaO | Na$_2$O | K$_2$O | Total |
|---|---|---|---|---|---|---|---|---|---|---|
| A1 | 51.98 | 1.59 | 19.44 | 7.00 | 0.06 | 3.61 | 7.75 | 5.93 | 2.65 | 100.00 |
| A2 | 51.24 | 1.52 | 19.45 | 7.14 | 0.22 | 3.56 | 7.98 | 6.36 | 2.54 | 100.00 |
| A3 | 50.96 | 1.60 | 19.44 | 7.56 | 0.20 | 3.60 | 8.27 | 5.87 | 2.51 | 100.00 |
| A4 | 52.11 | 1.50 | 19.14 | 7.42 | 0.12 | 3.76 | 7.59 | 5.67 | 2.69 | 100.00 |
| A5 | 52.04 | 1.57 | 19.73 | 7.04 | 0.22 | 3.66 | 7.67 | 5.59 | 2.47 | 100.00 |
| **average** | **51.67** | **1.56** | **19.44** | **7.23** | **0.16** | **3.64** | **7.85** | **5.88** | **2.57** | **100.00** |

*300_S4*
Experimental conditions:  0.9 XCO$_2^{fl}$; P=300 Mpa; T 1100°C

| spots | SiO$_2$ | TiO$_2$ | Al$_2$O$_3$ | FeO | MnO | MgO | CaO | Na$_2$O | K$_2$O | Total |
|---|---|---|---|---|---|---|---|---|---|---|
| A1 | 52.36 | 1.39 | 19.92 | 7.15 | 0.04 | 2.90 | 5.81 | 7.37 | 3.06 | 100.00 |
| A2 | 51.76 | 1.38 | 20.04 | 7.60 | 0.12 | 2.85 | 6.35 | 6.84 | 3.06 | 100.00 |
| A3 | 52.65 | 1.39 | 19.84 | 7.21 | 0.12 | 2.79 | 6.16 | 6.62 | 3.23 | 100.00 |
| A4 | 51.77 | 1.42 | 20.31 | 7.08 | 0.35 | 2.91 | 6.29 | 6.90 | 2.96 | 100.00 |
| A5 | 52.20 | 1.40 | 20.30 | 7.58 | 0.22 | 2.87 | 5.94 | 6.41 | 3.08 | 100.00 |
| **average** | **52.15** | **1.39** | **20.08** | **7.32** | **0.17** | **2.87** | **6.11** | **6.83** | **3.08** | **100.00** |

*100_S1*
Experimental conditions:  0.2 XCO$_2^{fl}$; P=100 Mpa; T 1100°C

| spots | SiO$_2$ | TiO$_2$ | Al$_2$O$_3$ | FeO | MnO | MgO | CaO | Na$_2$O | K$_2$O | Total |
|---|---|---|---|---|---|---|---|---|---|---|
| A1 | 50.29 | 1.83 | 18.50 | 7.43 | 0.26 | 5.16 | 9.95 | 4.25 | 2.34 | 100.00 |
| A2 | 50.74 | 1.79 | 18.25 | 7.73 | 0.11 | 4.96 | 10.17 | 3.99 | 2.26 | 100.00 |
| A3 | 50.47 | 1.77 | 18.40 | 7.75 | 0.29 | 4.87 | 9.81 | 4.37 | 2.26 | 100.00 |
| A4 | 50.50 | 1.78 | 18.24 | 7.71 | 0.16 | 4.88 | 10.01 | 4.43 | 2.30 | 100.00 |
| A5 | 50.82 | 1.76 | 18.47 | 7.05 | 0.15 | 5.06 | 10.11 | 4.23 | 2.35 | 100.00 |
| **average** | **50.57** | **1.78** | **18.37** | **7.53** | **0.19** | **4.99** | **10.01** | **4.25** | **2.30** | **100.00** |

*100_S2*
Experimental conditions:  0.4 XCO$_2^{fl}$; P=100 Mpa; T 1100°C

| spots | SiO$_2$ | TiO$_2$ | Al$_2$O$_3$ | FeO | MnO | MgO | CaO | Na$_2$O | K$_2$O | Total |
|---|---|---|---|---|---|---|---|---|---|---|
| A1 | 53.20 | 1.78 | 18.43 | 6.96 | 0.25 | 4.12 | 6.96 | 5.36 | 2.95 | 100.00 |
| A2 | 52.81 | 1.81 | 18.63 | 6.91 | 0.26 | 4.24 | 6.90 | 5.56 | 2.88 | 100.00 |
| A3 | 52.86 | 1.93 | 18.22 | 7.04 | 0.26 | 4.42 | 6.73 | 5.52 | 3.03 | 100.00 |
| A4 | 52.91 | 1.80 | 18.23 | 7.13 | 0.14 | 4.45 | 7.05 | 5.25 | 3.04 | 100.00 |
| **average** | **52.95** | **1.83** | **18.38** | **7.01** | **0.23** | **4.31** | **6.91** | **5.42** | **2.97** | **100.00** |

*100_S4*
Experimental conditions:  0.9 XCO$_2^{fl}$; P=100 Mpa; T 1100°C

| spots | SiO$_2$ | TiO$_2$ | Al$_2$O$_3$ | FeO | MnO | MgO | CaO | Na$_2$O | K$_2$O | Total |
|---|---|---|---|---|---|---|---|---|---|---|
| A1 | 52.32 | 1.71 | 18.71 | 7.69 | 0.14 | 3.53 | 6.40 | 5.88 | 3.62 | 100.00 |
| A2 | 51.47 | 1.71 | 19.09 | 7.80 | 0.17 | 3.60 | 6.38 | 6.13 | 3.65 | 100.00 |
| A3 | 51.95 | 1.56 | 19.08 | 7.66 | 0.26 | 3.54 | 6.18 | 6.13 | 3.64 | 100.00 |
| A4 | 51.83 | 1.66 | 18.85 | 7.69 | 0.28 | 3.42 | 6.43 | 6.10 | 3.74 | 100.00 |
| A5 | 51.87 | 1.57 | 18.99 | 7.39 | 0.17 | 3.58 | 6.30 | 6.45 | 3.70 | 100.00 |
| A6 | 51.99 | 1.58 | 18.92 | 7.75 | 0.32 | 3.39 | 6.10 | 6.28 | 3.67 | 100.00 |
| **average** | **51.90** | **1.63** | **18.94** | **7.66** | **0.22** | **3.51** | **6.30** | **6.16** | **3.67** | **100.00** |

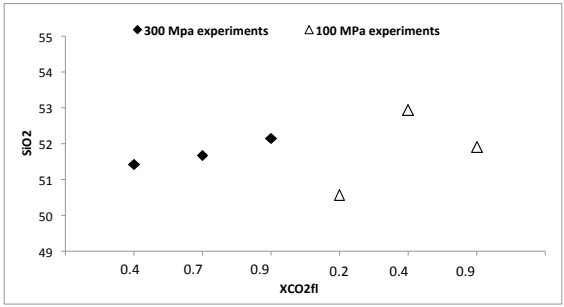
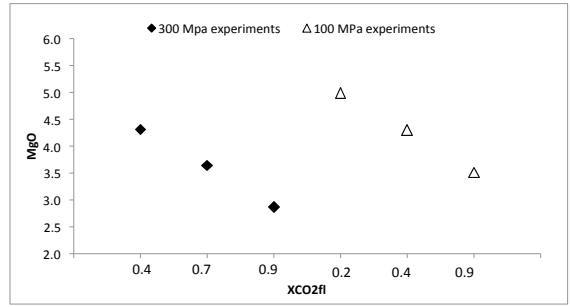
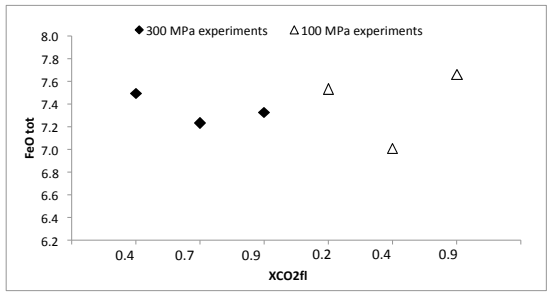
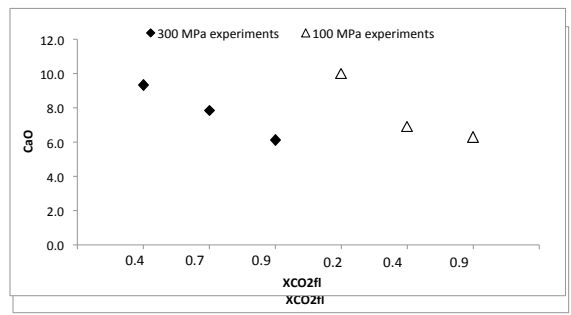
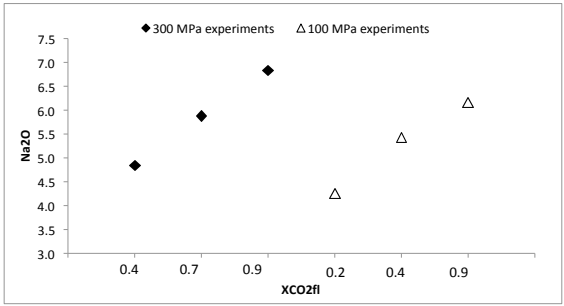
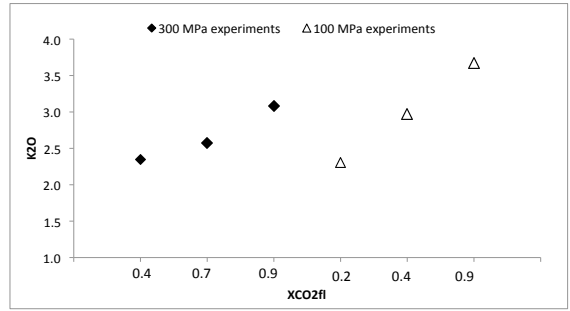
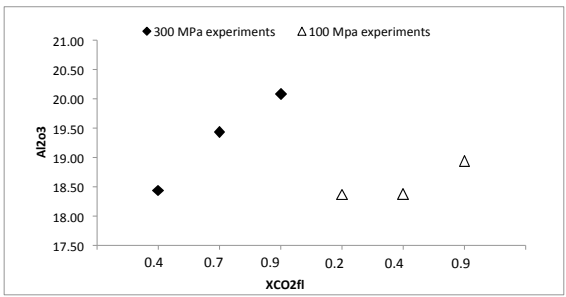
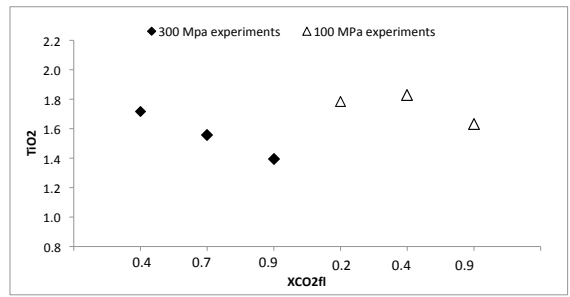

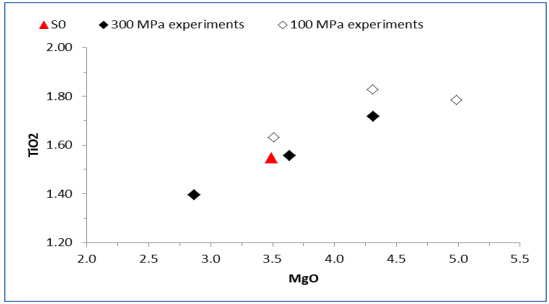
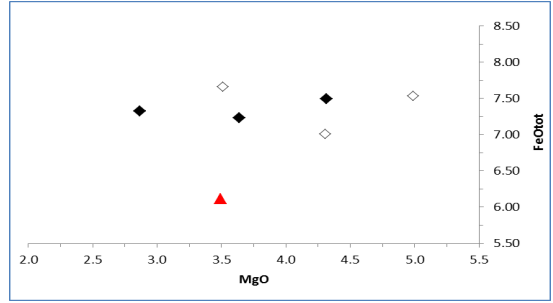
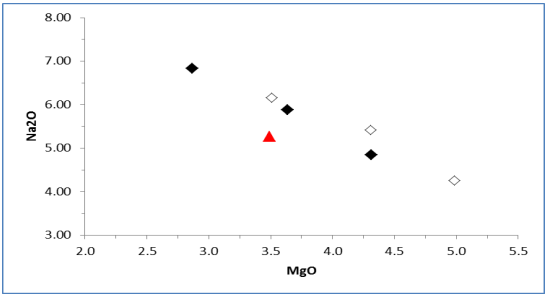
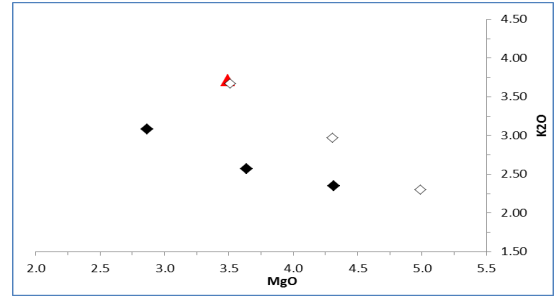
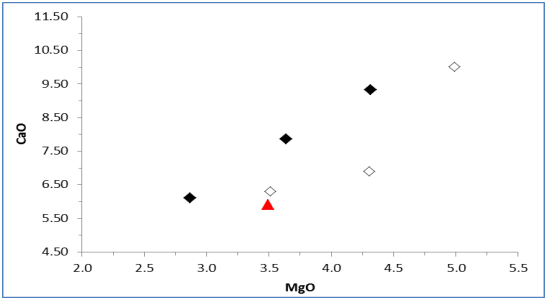
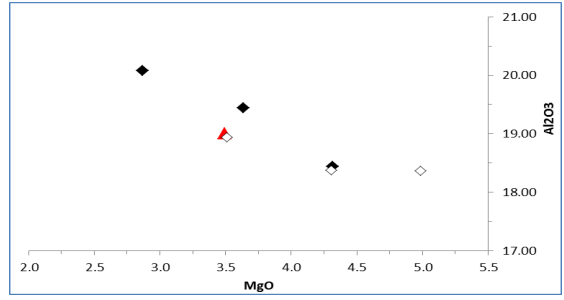

Mineral assemblage and compositions of the experimental samples
Px=pyroxene; Plg=plagioclase; Ox=opaque oxides.

## H2O-bearing system

### S0
Experimental conditions: 1 wt.% H2O; Thermal Cycling  76h  1080±20 °C ;  P=300 MPa.

| spots | SiO2 | TiO2 | Al2O3 | FeO | MnO | MgO | CaO | Na2O | K2O | Total | Diopside | Hedenbergit | Jadeite | Opx | Tschermakite | Mg# | Wo | En | Fs |
|---|---|---|---|---|---|---|---|---|---|---|---|---|---|---|---|---|---|---|---|
| Px1_S0_A1 | 41.36 | 2.73 | 10.94 | 11.63 | 0.07 | 9.51 | 21.42 | 0.53 | 0.00 | 98.13 | 0.48 | 0.14 | 0.00 | 0.08 | 0.33 | 0.77 | 48.98 | 30.27 | 20.75 |
| Px1_Sec0_A2 | 45.03 | 1.43 | 8.06 | 9.58 | 0.15 | 11.89 | 21.67 | 0.43 | 0.00 | 98.10 | 0.56 | 0.11 | 0.00 | 0.09 | 0.25 | 0.84 | 47.42 | 36.22 | 16.36 |
| Px1_Sec0_A3 | 42.74 | 2.09 | 10.89 | 10.03 | 0.21 | 10.27 | 21.74 | 0.63 | 0.00 | 98.40 | 0.51 | 0.10 | 0.00 | 0.07 | 0.32 | 0.83 | 49.56 | 32.59 | 17.86 |
| Px2_Sec0_A1 | 45.10 | 1.76 | 8.73 | 8.73 | 0.15 | 11.88 | 21.40 | 0.56 | 0.00 | 98.15 | 0.56 | 0.11 | 0.00 | 0.09 | 0.25 | 0.84 | 47.82 | 36.95 | 15.22 |
| Px2_Sec0_A2 | 44.69 | 2.01 | 9.12 | 8.69 | 0.14 | 11.35 | 21.71 | 0.60 | 0.01 | 98.18 | 0.56 | 0.12 | 0.00 | 0.08 | 0.25 | 0.83 | 49.03 | 35.66 | 15.31 |
| Px2_Sec0_A3 | 42.91 | 1.82 | 10.60 | 10.47 | 0.16 | 10.39 | 21.20 | 0.51 | 0.00 | 97.91 | 0.48 | 0.12 | 0.00 | 0.10 | 0.31 | 0.80 | 48.37 | 32.99 | 18.64 |
| Px3_Sec0_A1 | 42.38 | 1.91 | 11.14 | 11.18 | 0.18 | 10.30 | 20.48 | 0.64 | 0.01 | 98.03 | 0.44 | 0.11 | 0.00 | 0.12 | 0.33 | 0.80 | 47.05 | 32.91 | 20.04 |
| Px3_Sec0_A2 | 41.23 | 2.46 | 11.76 | 10.98 | 0.18 | 9.67 | 21.50 | 0.58 | 0.02 | 98.22 | 0.48 | 0.10 | 0.00 | 0.08 | 0.36 | 0.82 | 49.39 | 30.91 | 19.69 |
| Px4_Sec0_A1 | 43.87 | 2.04 | 10.20 | 9.39 | 0.29 | 11.09 | 20.76 | 0.62 | 0.00 | 97.97 | 0.49 | 0.12 | 0.00 | 0.11 | 0.28 | 0.81 | 47.70 | 35.45 | 16.85 |
| Px4_Sec0_A2 | 41.12 | 2.26 | 12.12 | 11.03 | 0.21 | 9.48 | 21.66 | 0.57 | 0.01 | 98.26 | 0.47 | 0.10 | 0.00 | 0.07 | 0.37 | 0.82 | 49.84 | 30.34 | 19.82 |
| Px4_Sec0_A3 | 46.12 | 1.66 | 7.86 | 8.45 | 0.16 | 12.53 | 21.38 | 0.61 | 0.02 | 98.63 | 0.58 | 0.10 | 0.00 | 0.10 | 0.22 | 0.85 | 47.08 | 38.39 | 14.53 |
| Px4_Sec0_A4 | 43.42 | 1.68 | 10.18 | 9.97 | 0.14 | 10.71 | 21.62 | 0.49 | 0.00 | 98.08 | 0.51 | 0.11 | 0.00 | 0.08 | 0.30 | 0.82 | 48.79 | 33.64 | 17.56 |
| Px4_Sec0_A5 | 44.02 | 2.14 | 9.53 | 9.53 | 0.16 | 10.71 | 21.18 | 0.59 | 0.02 | 98.78 | 0.53 | 0.09 | 0.00 | 0.10 | 0.28 | 0.86 | 47.06 | 36.41 | 16.53 |
| Px5_Sec0_A1 | 43.64 | 2.46 | 9.39 | 9.50 | 0.25 | 11.63 | 21.03 | 0.77 | 0.03 | 98.45 | 0.55 | 0.07 | 0.00 | 0.09 | 0.29 | 0.88 | 47.13 | 36.25 | 16.61 |
| Px5_Sec0_A2 | 41.89 | 2.01 | 11.64 | 11.05 | 0.19 | 9.89 | 21.43 | 0.63 | 0.02 | 98.55 | 0.47 | 0.10 | 0.00 | 0.08 | 0.35 | 0.83 | 48.91 | 31.41 | 19.68 |
| Px5_Sec0_A3 | 44.96 | 2.11 | 8.84 | 9.22 | 0.15 | 11.87 | 21.05 | 0.62 | 0.02 | 98.68 | 0.54 | 0.11 | 0.00 | 0.11 | 0.25 | 0.83 | 47.01 | 36.91 | 16.08 |

| Plagioclase | SiO2 | TiO2 | Al2O3 | FeO | MnO | MgO | CaO | Na2O | K2O | Total | An% | Ab% | Or% |
|---|---|---|---|---|---|---|---|---|---|---|---|---|---|
| Plg1_exp1_Sec0_A1 | 51.77 | 0.07 | 29.78 | 0.86 | 0.05 | 0.08 | 13.22 | 3.84 | 0.40 | 100.06 | 64.00 | 33.67 | 2.33 |
| Plg1_exp1_Sec0_A2 | 53.93 | 0.12 | 27.98 | 0.94 | 0.02 | 0.07 | 11.38 | 4.85 | 0.54 | 99.83 | 54.72 | 42.20 | 3.08 |
| Plg1_exp1_Sec0_A3 | 52.76 | 0.06 | 28.75 | 1.04 | 0.00 | 0.12 | 12.20 | 4.29 | 0.49 | 99.71 | 59.35 | 37.79 | 2.87 |
| Plg1_exp1_Sec0_A4 | 52.35 | 0.09 | 28.74 | 0.99 | 0.03 | 0.13 | 12.36 | 4.44 | 0.47 | 99.59 | 59.02 | 38.33 | 2.65 |
| Plg1_exp1_Sec0_A5 | 52.98 | 0.09 | 28.44 | 0.95 | 0.02 | 0.08 | 12.05 | 4.37 | 0.51 | 99.49 | 58.58 | 38.46 | 2.96 |
| Plg2_exp1_Sec0_A1 | 52.93 | 0.07 | 29.04 | 1.04 | -0.05 | 0.08 | 11.79 | 4.36 | 0.48 | 99.73 | 58.22 | 38.96 | 2.82 |
| Plg2_exp1_Sec0_A2 | 52.69 | 0.08 | 28.68 | 1.09 | 0.01 | 0.15 | 11.76 | 4.32 | 0.47 | 99.25 | 58.40 | 38.80 | 2.81 |
| Plg3_exp1_Sec0_A1 | 52.72 | 0.09 | 29.02 | 0.92 | -0.06 | 0.10 | 11.95 | 4.30 | 0.48 | 99.52 | 58.84 | 38.32 | 2.83 |
| Plg3_exp1_Sec0_A2 | 52.43 | 0.09 | 28.99 | 0.75 | 0.02 | 0.12 | 12.31 | 4.17 | 0.45 | 99.32 | 60.35 | 37.03 | 2.62 |
| Plg3_exp1_Sec0_A3 | 52.72 | 0.10 | 29.23 | 0.74 | -0.07 | 0.08 | 11.83 | 4.44 | 0.48 | 99.54 | 57.90 | 39.30 | 2.79 |
| Plg3_exp1_Sec0_A4 | 51.46 | 0.07 | 29.33 | 0.91 | 0.01 | 0.10 | 12.69 | 4.01 | 0.44 | 99.01 | 62.00 | 35.47 | 2.53 |
| Plg4_exp1_Sec0_A1 | 52.86 | 0.11 | 28.69 | 0.92 | 0.00 | 0.08 | 11.83 | 4.38 | 0.45 | 99.32 | 58.31 | 39.07 | 2.62 |
| Plg4_exp1_Sec0_A2 | 53.14 | 0.09 | 28.37 | 1.05 | 0.05 | 0.10 | 11.42 | 4.54 | 0.48 | 99.25 | 56.51 | 40.66 | 2.83 |
| Plg4_exp1_Sec0_A3 | 53.42 | 0.11 | 28.18 | 1.02 | -0.03 | 0.11 | 11.23 | 4.66 | 0.48 | 99.19 | 55.48 | 41.67 | 2.85 |
| Plg4_exp1_Sec0_A4 | 52.36 | 0.13 | 28.91 | 1.08 | -0.06 | 0.07 | 11.66 | 4.41 | 0.47 | 99.04 | 57.71 | 39.52 | 2.78 |

| Oxides | SiO2 | TiO2 | Al2O3 | FeO | MnO | MgO | CaO | Na2O | K2O | Total |
|---|---|---|---|---|---|---|---|---|---|---|
| Ox1 | 0.09 | 6.06 | 9.71 | 69.57 | 0.48 | 7.79 | 0.07 | 0.02 | 0.02 | 93.75 |
| Ox1 | 0.13 | 6.26 | 9.94 | 70.49 | 0.46 | 7.75 | 0.06 | 0.04 | 0.00 | 95.12 |
| Ox1 | 0.11 | 6.27 | 9.73 | 69.25 | 0.57 | 7.63 | 0.05 | 0.02 | 0.01 | 93.63 |
| Ox1 | 0.14 | 6.14 | 9.71 | 69.91 | 0.51 | 7.70 | 0.04 | 0.04 | 0.01 | 94.21 |
| Ox2 | 0.09 | 6.25 | 9.26 | 70.49 | 0.54 | 7.62 | 0.08 | -0.01 | 0.03 | 94.34 |
| Ox2 | 0.11 | 6.28 | 9.17 | 70.46 | 0.50 | 7.66 | 0.09 | 0.05 | 0.02 | 94.34 |
| Ox2 | 0.13 | 6.34 | 9.16 | 71.24 | 0.36 | 7.61 | 0.12 | 0.00 | 0.02 | 94.99 |
| Ox2 | 0.10 | 6.11 | 9.57 | 69.30 | 0.47 | 7.47 | 0.10 | 0.07 | 0.03 | 93.23 |
| Ox3 | 0.14 | 6.41 | 9.56 | 70.36 | 0.42 | 7.63 | 0.10 | 0.01 | 0.04 | 94.68 |
| Ox3 | 0.12 | 6.55 | 9.38 | 70.74 | 0.51 | 7.66 | 0.03 | 0.04 | 0.02 | 95.06 |
| Ox3 | 0.11 | 6.52 | 9.56 | 69.95 | 0.54 | 7.73 | 0.06 | 0.00 | 0.02 | 94.48 |
| Ox3 | 0.09 | 6.56 | 9.34 | 70.15 | 0.57 | 7.43 | 0.14 | 0.04 | 0.04 | 94.37 |

## H2O+CO2-bearing system

### 300_S2
Experimental conditions:  0.4 XCO2; P=300 Mpa; T 1100°C

| spots | SiO2 | TiO2 | Al2O3 | FeO | MnO | MgO | CaO | Na2O | K2O | Total | Diopside | Hedenbergit | Jadeite | Opx | Tschermakite | Mg# | Wo | En | Fs |
|---|---|---|---|---|---|---|---|---|---|---|---|---|---|---|---|---|---|---|---|
| Px1_Ex1_Sec2_A31 | 44.84 | 1.88 | 9.19 | 9.09 | 0.15 | 11.79 | 21.54 | 0.55 | 0.01 | 98.88 | 0.55 | 0.10 | 0.00 | 0.09 | 0.26 | 0.84 | 47.82 | 36.42 | 15.76 |
| Px1_Ex1_Sec2_A32 | 43.05 | 1.94 | 10.43 | 10.54 | 0.10 | 10.63 | 21.81 | 0.51 | 0.01 | 98.92 | 0.51 | 0.10 | 0.00 | 0.08 | 0.32 | 0.83 | 48.66 | 32.99 | 18.35 |
| Px1_Ex1_Sec2_A33 | 45.05 | 1.77 | 9.03 | 9.69 | 0.21 | 11.69 | 21.60 | 0.50 | 0.01 | 99.34 | 0.53 | 0.11 | 0.00 | 0.10 | 0.26 | 0.83 | 47.56 | 35.79 | 16.65 |
| Px1_Ex1_Sec2_A34 | 45.93 | 1.69 | 8.02 | 8.68 | 0.23 | 12.40 | 21.78 | 0.63 | 0.02 | 99.15 | 0.59 | 0.09 | 0.00 | 0.08 | 0.24 | 0.87 | 47.55 | 37.66 | 14.79 |
| Px1_Ex1_Sec2_A35 | 44.08 | 1.76 | 9.66 | 9.75 | 0.24 | 11.20 | 21.91 | 0.54 | 0.03 | 98.92 | 0.54 | 0.09 | 0.00 | 0.08 | 0.29 | 0.86 | 48.59 | 34.55 | 16.87 |
| Px2_Ex1_Sec2_A1 | 45.11 | 1.85 | 8.56 | 9.17 | 0.19 | 11.99 | 22.78 | 0.58 | 0.02 | 100.07 | 0.61 | 0.06 | 0.00 | 0.05 | 0.28 | 0.91 | 48.87 | 35.77 | 15.36 |
| Px2_Ex1_Sec2_A2 | 41.44 | 2.62 | 11.73 | 11.52 | 0.14 | 9.96 | 21.14 | 0.60 | 0.00 | 99.02 | 0.46 | 0.11 | 0.00 | 0.10 | 0.36 | 0.81 | 48.06 | 31.51 | 20.43 |
| Px2_Ex1_Sec2_A3 | 44.07 | 1.80 | 10.53 | 10.30 | 0.13 | 10.98 | 21.07 | 0.51 | 0.02 | 99.28 | 0.47 | 0.12 | 0.00 | 0.11 | 0.29 | 0.79 | 47.47 | 34.43 | 18.10 |
| Px2_Ex1_Sec2_A4 | 45.43 | 1.86 | 8.80 | 9.11 | 0.25 | 12.18 | 21.20 | 0.58 | 0.01 | 99.17 | 0.54 | 0.10 | 0.00 | 0.11 | 0.25 | 0.84 | 46.84 | 37.45 | 15.71 |
| Px2_Ex1_Sec2_A6 | 43.88 | 1.98 | 9.40 | 9.06 | 0.18 | 11.21 | 22.41 | 0.58 | 0.04 | 98.56 | 0.59 | 0.07 | 0.00 | 0.05 | 0.29 | 0.89 | 49.70 | 34.61 | 15.68 |

| oxide | SiO2 | TiO2 | Al2O3 | FeO | MnO | MgO | CaO | Na2O | K2O | Total |
|---|---|---|---|---|---|---|---|---|---|---|
| Ox1 | 0.11 | 4.19 | 8.70 | 75.94 | 0.32 | 7.27 | 0.15 | 0.06 | 0.03 | 96.65 |
| Ox1 | 0.09 | 4.20 | 8.51 | 74.93 | 0.42 | 7.29 | 0.11 | 0.01 | 0.02 | 95.57 |
| Ox1 | 0.08 | 4.23 | 8.49 | 76.74 | 0.46 | 7.31 | 0.13 | 0.05 | 0.00 | 97.50 |
| Ox2 | 0.07 | 4.20 | 8.52 | 77.03 | 0.26 | 7.49 | 0.12 | 0.00 | 0.01 | 97.71 |
| Ox2 | 0.08 | 4.19 | 8.65 | 77.07 | 0.43 | 7.36 | 0.11 | 0.01 | 0.02 | 97.92 |

### 300_S3
Experimental conditions:  0.9 XCO2; P=300 Mpa; T 1100°C

| spots | SiO2 | TiO2 | Al2O3 | FeO | MnO | MgO | CaO | Na2O | K2O | Total | Diopside | Hedenbergit | Jadeite | Opx | Tschermakite | Mg# | Wo | En | Fs |
|---|---|---|---|---|---|---|---|---|---|---|---|---|---|---|---|---|---|---|---|
| Px1_Ex1_Sec1_A1 | 44.68 | 2.04 | 10.36 | 9.70 | 0.32 | 10.89 | 20.44 | 0.60 | 0.09 | 98.78 | 0.46 | 0.15 | 0.00 | 0.13 | 0.26 | 0.76 | 47.36 | 35.11 | 17.54 |
| Px1_Ex1_Sec1_A2 | 42.56 | 1.87 | 11.32 | 11.17 | 0.20 | 10.24 | 21.21 | 0.61 | 0.02 | 99.00 | 0.46 | 0.10 | 0.00 | 0.10 | 0.34 | 0.82 | 48.01 | 32.26 | 19.73 |
| Px1_Ex1_Sec1_A3 | 42.41 | 1.84 | 11.62 | 10.53 | 0.25 | 10.18 | 21.39 | 0.54 | 0.01 | 98.53 | 0.47 | 0.10 | 0.00 | 0.09 | 0.34 | 0.82 | 48.86 | 32.36 | 18.78 |
| Px1_Ex1_Sec1_A4 | 45.96 | 1.62 | 8.29 | 8.88 | 0.19 | 11.83 | 21.82 | 0.60 | 0.02 | 99.03 | 0.57 | 0.12 | 0.00 | 0.08 | 0.23 | 0.83 | 48.26 | 36.41 | 15.33 |
| Px2_Ex1_Sec1_A1 | 45.73 | 1.86 | 7.93 | 9.32 | 0.16 | 11.93 | 20.86 | 0.65 | 0.02 | 98.29 | 0.54 | 0.13 | 0.00 | 0.11 | 0.22 | 0.81 | 46.63 | 37.12 | 16.26 |
| Px2_Ex1_Sec1_A2 | 44.29 | 1.97 | 10.08 | 9.81 | 0.18 | 11.16 | 21.26 | 0.61 | 0.02 | 99.20 | 0.51 | 0.11 | 0.00 | 0.10 | 0.28 | 0.82 | 47.83 | 34.94 | 17.22 |
| Px2_Ex1_Sec1_A3 | 43.64 | 1.88 | 10.74 | 10.99 | 0.15 | 10.53 | 20.28 | 0.58 | 0.01 | 98.64 | 0.43 | 0.14 | 0.00 | 0.13 | 0.29 | 0.75 | 46.60 | 33.68 | 19.72 |
| Px2_Ex1_Sec1_A4 | 45.43 | 1.84 | 8.33 | 8.91 | 0.20 | 11.31 | 22.81 | 0.70 | 0.03 | 99.35 | 0.62 | 0.09 | 0.00 | 0.03 | 0.25 | 0.87 | 50.13 | 34.59 | 15.28 |
| Px3_Ex1_Sec1_A1 | 44.05 | 1.78 | 8.87 | 10.78 | 0.28 | 11.88 | 20.25 | 0.45 | 0.00 | 98.07 | 0.48 | 0.11 | 0.00 | 0.14 | 0.28 | 0.82 | 44.81 | 36.58 | 18.61 |
| Px3_Ex1_Sec1_A3 | 43.66 | 2.15 | 10.44 | 9.89 | 0.19 | 10.82 | 21.31 | 0.56 | 0.01 | 98.84 | 0.50 | 0.12 | 0.00 | 0.10 | 0.29 | 0.80 | 48.33 | 34.15 | 17.51 |

| oxides | SiO2 | TiO2 | Al2O3 | FeO | MnO | MgO | CaO | Na2O | K2O | Total |
|---|---|---|---|---|---|---|---|---|---|---|
| Ox1 | 0.12 | 4.79 | 9.55 | 75.39 | 0.47 | 7.20 | 0.11 | 0.04 | 0.02 | 97.69 |
| Ox1 | 0.05 | 4.79 | 9.50 | 75.13 | 0.42 | 7.14 | 0.12 | 0.01 | 0.02 | 97.20 |
| Ox1 | 0.09 | 4.72 | 9.68 | 74.94 | 0.44 | 6.89 | 0.15 | 0.02 | 0.03 | 96.96 |
| Ox2 | 0.09 | 4.75 | 9.69 | 74.73 | 0.39 | 7.19 | 0.16 | 0.02 | 0.02 | 97.02 |
| Ox2 | 0.08 | 4.78 | 9.69 | 74.61 | 0.45 | 7.05 | 0.06 | 0.01 | 0.01 | 96.72 |
| Ox2 | 0.08 | 4.77 | 9.47 | 74.98 | 0.46 | 7.03 | 0.17 | 0.01 | 0.02 | 96.96 |
| Ox2 | 0.09 | 4.69 | 9.43 | 74.67 | 0.48 | 7.14 | 0.12 | 0.06 | 0.04 | 96.60 |

### 300_S4
Experimental conditions:  0.9 XCO2; P=300 Mpa; T 1100°C

| spots | SiO2 | TiO2 | Al2O3 | FeO | MnO | MgO | CaO | Na2O | K2O | Total | Diopside | Hedenbergit | Jadeite | Opx | Tschermakite | Mg# | Wo | En | Fs |
|---|---|---|---|---|---|---|---|---|---|---|---|---|---|---|---|---|---|---|---|
| Px1_Ex1_Sec4_A3 | 41.35 | 2.53 | 12.10 | 11.94 | 0.14 | 9.39 | 20.83 | 0.58 | 0.01 | 98.73 | 0.43 | 0.11 | 0.00 | 0.11 | 0.35 | 0.75 | 48.21 | 30.23 | 21.57 |
| Px1_Ex1_Sec4_A4 | 44.71 | 2.33 | 10.07 | 8.99 | 0.22 | 11.20 | 21.27 | 0.67 | 0.01 | 99.26 | 0.52 | 0.13 | 0.00 | 0.10 | 0.26 | 0.80 | 48.47 | 35.53 | 16.00 |
| Px1_Ex1_Sec4_A5 | 48.76 | 1.60 | 5.93 | 8.15 | 0.22 | 12.47 | 21.92 | 0.90 | 0.03 | 99.75 | 0.62 | 0.14 | 0.00 | 0.06 | 0.15 | 0.82 | 48.03 | 38.03 | 13.94 |
| Px2_Ex1_Sec4_A3 | 41.37 | 2.18 | 11.57 | 11.72 | 0.15 | 9.78 | 20.87 | 0.54 | 0.00 | 98.04 | 0.44 | 0.12 | 0.00 | 0.10 | 0.35 | 0.79 | 47.84 | 31.18 | 20.97 |
| Px2_Ex1_Sec4_A4 | 40.54 | 2.65 | 10.82 | 12.54 | 0.26 | 10.07 | 20.11 | 0.60 | 0.01 | 97.35 | 0.44 | 0.10 | 0.00 | 0.12 | 0.36 | 0.82 | 45.81 | 31.90 | 22.30 |
| Px2_Ex1_Sec4_A4 | 44.66 | 1.92 | 8.59 | 9.82 | 0.23 | 12.30 | 21.19 | 0.57 | 0.02 | 99.09 | 0.55 | 0.08 | 0.00 | 0.11 | 0.27 | 0.88 | 46.09 | 37.23 | 16.68 |
| Px2_Ex1_Sec4_A5 | 43.77 | 2.35 | 9.55 | 9.62 | 0.31 | 11.98 | 21.14 | 0.65 | 0.01 | 99.07 | 0.54 | 0.07 | 0.00 | 0.10 | 0.30 | 0.89 | 46.65 | 36.78 | 16.57 |
| Px2_Ex1_Sec4_A6 | 46.46 | 1.74 | 7.03 | 8.53 | 0.22 | 13.00 | 21.70 | 0.67 | 0.03 | 99.16 | 0.61 | 0.07 | 0.00 | 0.08 | 0.22 | 0.90 | 46.71 | 38.95 | 14.34 |
| Px2_Ex1_Sec4_A7 | 44.51 | 2.46 | 9.52 | 9.66 | 0.17 | 10.97 | 21.07 | 1.05 | 0.13 | 99.37 | 0.54 | 0.09 | 0.00 | 0.07 | 0.27 | 0.86 | 48.02 | 34.79 | 17.19 |
| Px3_Ex1_Sec4_A1 | 45.33 | 2.17 | 8.89 | 9.14 | 0.27 | 11.60 | 20.93 | 0.68 | 0.03 | 98.77 | 0.53 | 0.13 | 0.00 | 0.11 | 0.23 | 0.80 | 47.54 | 36.51 | 16.14 |
| Px3_Ex1_Sec4_A2 | 41.34 | 2.88 | 11.90 | 12.65 | 0.26 | 9.33 | 20.60 | 0.67 | 0.06 | 99.42 | 0.42 | 0.13 | 0.00 | 0.11 | 0.35 | 0.75 | 47.40 | 29.88 | 22.72 |
| Px3_Ex1_Sec4_A3 | 41.86 | 2.18 | 11.61 | 11.86 | 0.28 | 9.91 | 20.96 | 0.54 | 0.01 | 98.95 | 0.44 | 0.11 | 0.00 | 0.11 | 0.35 | 0.79 | 47.72 | 31.34 | 21.04 |
| Px3_Ex1_Sec4_A5 | 44.40 | 2.29 | 9.38 | 9.35 | 0.24 | 11.52 | 21.44 | 0.58 | 0.01 | 98.98 | 0.54 | 0.11 | 0.00 | 0.09 | 0.27 | 0.83 | 47.89 | 35.81 | 16.30 |
| Px3_Ex1_Sec4_A6 | 44.49 | 2.32 | 10.04 | 9.18 | 0.21 | 11.03 | 21.37 | 0.64 | 0.02 | 99.06 | 0.51 | 0.12 | 0.00 | 0.09 | 0.26 | 0.82 | 48.77 | 34.97 | 16.32 |
| Px4_Ex1_Sec4_A8 | 45.50 | 1.87 | 6.87 | 10.32 | 0.13 | 11.23 | 21.61 | 1.24 | 0.04 | 98.66 | 0.61 | 0.07 | 0.00 | 0.04 | 0.23 | 0.89 | 47.71 | 34.49 | 17.79 |
| Px4_Ex1_Sec4_A9 | 45.85 | 1.99 | 8.99 | 8.67 | 0.27 | 11.17 | 21.61 | 0.88 | 0.12 | 99.28 | 0.55 | 0.11 | 0.00 | 0.07 | 0.23 | 0.83 | 49.20 | 35.40 | 15.40 |
| Px4_Ex1_Sec4_A10 | 44.46 | 2.14 | 8.57 | 9.59 | 0.23 | 11.69 | 20.91 | 0.71 | 0.02 | 98.10 | 0.54 | 0.10 | 0.00 | 0.10 | 0.26 | 0.83 | 46.82 | 36.41 | 16.77 |

| Plagioclase | SiO2 | TiO2 | Al2O3 | FeO | MnO | MgO | CaO | Na2O | K2O | Total | An% | Ab% | Or% |
|---|---|---|---|---|---|---|---|---|---|---|---|---|---|
| Plg1_exp1_A1 | 53.21 | 0.12 | 28.67 | 1.19 | 0.05 | 0.10 | 11.67 | 4.34 | 0.41 | 99.75 | 58.31 | 39.27 | 2.42 |
| Plg1_exp1_A2 | 53.16 | 0.09 | 29.07 | 1.16 | 0.02 | 0.07 | 11.64 | 4.32 | 0.46 | 100.01 | 58.18 | 39.11 | 2.72 |
| Plg1_exp1_A3 | 52.15 | 0.10 | 29.24 | 1.10 | 0.01 | 0.12 | 12.26 | 4.28 | 0.37 | 99.63 | 59.93 | 37.91 | 2.16 |
| Plg1_exp1_A4 | 53.29 | 0.11 | 28.94 | 1.02 | -0.01 | 0.12 | 11.53 | 4.48 | 0.47 | 99.95 | 57.08 | 40.17 | 2.75 |
| Plg1_exp1_A5 | 52.82 | 0.10 | 29.29 | 0.96 | 0.03 | 0.06 | 12.17 | 4.17 | 0.40 | 99.98 | 60.29 | 37.37 | 2.34 |
| Plg1_exp1_A6 | 51.50 | 0.11 | 29.88 | 1.23 | 0.05 | 0.06 | 13.13 | 3.77 | 0.31 | 100.04 | 64.57 | 33.59 | 1.84 |
| Plg1_exp1_A7 | 52.71 | 0.09 | 29.11 | 0.86 | 0.05 | 0.07 | 11.99 | 4.37 | 0.41 | 99.65 | 58.82 | 38.78 | 2.40 |
| Plg1_exp1_A8 | 52.34 | 0.07 | 29.08 | 1.00 | 0.05 | 0.12 | 12.40 | 4.15 | 0.41 | 99.61 | 60.82 | 36.80 | 2.38 |
| Plg1_exp1_A9 | 53.04 | 0.09 | 28.85 | 0.82 | 0.00 | 0.05 | 11.51 | 4.49 | 0.40 | 99.24 | 57.26 | 40.40 | 2.34 |
| Plg1_exp1_A10 | 52.68 | 0.08 | 28.92 | 0.85 | 0.03 | 0.03 | 11.36 | 4.58 | 0.37 | 98.91 | 56.53 | 41.28 | 2.18 |
| Plg1_exp1_A11 | 53.05 | 0.10 | 29.27 | 1.16 | 0.00 | 0.07 | 11.74 | 4.36 | 0.41 | 100.15 | 58.38 | 39.18 | 2.44 |
| Plg1_exp1_A12 | 53.03 | 0.06 | 29.07 | 1.10 | 0.01 | 0.08 | 11.79 | 4.42 | 0.43 | 99.98 | 58.07 | 39.41 | 2.51 |
| Plg2_exp1_A1 | 52.35 | 0.07 | 28.99 | 1.02 | -0.03 | 0.12 | 12.36 | 4.08 | 0.39 | 99.11 | 58.67 | 36.56 | 2.27 |
| Plg2_exp1_A2 | 52.04 | 0.09 | 28.98 | 1.29 | 0.05 | 0.08 | 12.13 | 4.29 | 0.39 | 99.33 | 59.62 | 38.12 | 2.26 |
| Plg2_exp1_A3 | 52.17 | 0.11 | 29.07 | 0.97 | 0.00 | 0.09 | 11.97 | 4.32 | 0.36 | 99.07 | 59.21 | 38.69 | 2.10 |
| Plg2_exp1_A4 | 52.51 | 0.11 | 28.90 | 0.92 | 0.00 | 0.08 | 11.85 | 4.36 | 0.39 | 99.11 | 58.67 | 39.05 | 2.27 |
| Plg2_exp1_A5 | 52.39 | 0.09 | 28.99 | 1.09 | 0.02 | 0.09 | 12.13 | 4.29 | 0.39 | 99.25 | 58.03 | 39.86 | 2.11 |
| Plg2_exp1_A6 | 53.14 | 0.08 | 28.74 | 0.85 | 0.00 | 0.06 | 11.65 | 4.47 | 0.37 | 99.35 | 57.75 | 40.08 | 2.18 |
| Plg2_exp1_A7 | 52.87 | 0.09 | 29.21 | 0.95 | -0.02 | 0.07 | 11.64 | 4.37 | 0.36 | 99.55 | 58.26 | 39.58 | 2.16 |
| Plg3_exp1_A1 | 52.71 | 0.07 | 28.81 | 1.20 | -0.11 | 0.12 | 11.47 | 4.53 | 0.39 | 99.17 | 56.98 | 40.72 | 2.30 |
| Plg3_exp1_A2 | 52.50 | 0.08 | 28.58 | 1.07 | -0.01 | 0.08 | 11.83 | 4.20 | 0.39 | 98.74 | 59.48 | 38.19 | 2.33 |
| Plg3_exp1_A3 | 51.38 | 0.13 | 29.02 | 1.12 | -0.03 | 0.07 | 12.43 | 4.20 | 0.36 | 98.68 | 60.75 | 37.16 | 2.09 |
| Plg3_exp1_A4 | 51.90 | 0.08 | 28.92 | 1.07 | 0.03 | 0.06 | 12.38 | 4.32 | 0.31 | 99.09 | 60.16 | 38.02 | 1.82 |
| Plg3_exp1_A5 | 52.76 | 0.10 | 28.44 | 1.21 | 0.02 | 0.13 | 11.67 | 4.27 | 0.43 | 99.01 | 58.64 | 38.79 | 2.57 |
| Plg3_exp1_A6 | 51.94 | 0.12 | 28.77 | 1.18 | 0.02 | 0.10 | 11.99 | 4.51 | 0.37 | 99.01 | 58.22 | 39.62 | 2.16 |

| oxide | SiO2 | TiO2 | Al2O3 | FeO | MnO | MgO | CaO | Na2O | K2O | Total |
|---|---|---|---|---|---|---|---|---|---|---|
| Ox1 | 0.08 | 7.40 | 9.65 | 73.03 | 0.54 | 6.53 | 0.13 | 0.02 | 0.02 | 97.36 |
| Ox1 | 0.11 | 7.42 | 9.55 | 73.58 | 0.54 | 6.61 | 0.22 | 0.02 | 0.03 | 98.05 |
| Ox1 | 0.13 | 7.44 | 9.46 | 72.45 | 0.44 | 6.49 | 0.23 | 0.02 | 0.02 | 96.68 |
| Ox2 | 0.11 | 7.14 | 9.93 | 73.23 | 0.55 | 6.26 | 0.15 | 0.00 | 0.03 | 97.40 |
| Ox2 | 0.10 | 7.11 | 9.82 | 74.38 | 0.49 | 6.20 | 0.05 | 0.01 | 0.01 | 98.17 |
| Ox2 | 0.09 | 7.06 | 9.60 | 73.39 | 0.51 | 6.18 | 0.04 | 0.01 | -0.01 | 96.88 |
| Ox2 | 0.09 | 7.06 | 9.93 | 73.27 | 0.53 | 6.25 | 0.12 | 0.02 | 0.02 | 97.30 |

### 100_S1
Experimental conditions:  0.2 XCO2; P=100 Mpa; T 1100°C

| spots | SiO2 | TiO2 | Al2O3 | FeO | MnO | MgO | CaO | Na2O | K2O | Total | Diopside | Hedenbergit | Jadeite | Opx | Tschermakite | Mg# | Wo | En | Fs |
|---|---|---|---|---|---|---|---|---|---|---|---|---|---|---|---|---|---|---|---|
| Cpx1_Sec1_A1 | 42.80 | 2.37 | 9.86 | 9.94 | 0.11 | 11.61 | 22.23 | 0.45 | 0.00 | 99.36 | 0.58 | 0.05 | 0.00 | 0.04 | 0.33 | 0.92 | 48.17 | 35.01 | 16.82 |
| Cpx1_Sec1_A2 | 42.38 | 2.48 | 9.79 | 9.66 | 0.14 | 11.49 | 22.55 | 0.40 | 0.00 | 98.91 | 0.60 | 0.05 | 0.00 | 0.03 | 0.33 | 0.93 | 48.90 | 34.73 | 16.38 |
| Cpx1_Sec1_A3 | 41.87 | 2.78 | 10.18 | 9.88 | 0.09 | 11.19 | 22.38 | 0.44 | -0.01 | 98.81 | 0.59 | 0.05 | 0.00 | 0.06 | 0.34 | 0.92 | 49.01 | 34.10 | 16.89 |
| Cpx1_Sec1_A4 | 47.19 | 1.88 | 7.96 | 7.82 | 0.13 | 12.41 | 23.15 | 0.40 | 0.01 | 101.04 | 0.73 | 0.05 | 0.00 | 0.06 | 0.21 | 0.84 | 49.77 | 37.12 | 13.11 |
| Cpx1_Sec1_A5 | 45.60 | 1.53 | 6.82 | 9.04 | 0.15 | 13.44 | 22.50 | 0.39 | 0.02 | 99.54 | 0.89 | 0.04 | -0.26 | 0.07 | -0.04 | 0.96 | 46.63 | 38.75 | 14.62 |

### 100_S2
Experimental conditions:  0.4 XCO2; P=100 Mpa; T 1100°C

| spots | SiO2 | TiO2 | Al2O3 | FeO | MnO | MgO | CaO | Na2O | K2O | Total | Diopside | Hedenbergit | Jadeite | Opx | Tschermakite | Mg# | Wo | En | Fs |
|---|---|---|---|---|---|---|---|---|---|---|---|---|---|---|---|---|---|---|---|
| Cpx1_A1 | 47.60 | 1.57 | 5.36 | 8.51 | 0.19 | 14.45 | 21.08 | 0.40 | 0.02 | 99.18 | 0.82 | 0.03 | 0.00 | 0.13 | 0.18 | 0.85 | 44.08 | 42.04 | 13.88 |
| Cpx2_A1 | 43.28 | 2.08 | 10.13 | 9.52 | 0.23 | 11.21 | 21.43 | 0.48 | 0.02 | 98.37 | 0.52 | 0.09 | 0.00 | 0.09 | 0.30 | 0.85 | 48.17 | 35.13 | 16.70 |
| Cpx3_A1 | 43.83 | 2.10 | 10.46 | 9.22 | 0.25 | 10.91 | 21.73 | 0.44 | 0.03 | 99.01 | 0.52 | 0.12 | 0.00 | 0.08 | 0.29 | 0.81 | 49.26 | 34.43 | 16.31 |
| Cpx4_A1 | 43.52 | 2.53 | 9.51 | 9.02 | 0.17 | 12.01 | 20.82 | 0.60 | 0.01 | 98.18 | 0.55 | 0.08 | 0.00 | 0.11 | 0.28 | 0.87 | 46.72 | 37.48 | 15.79 |
| Cpx5_A1 | 42.44 | 2.39 | 11.42 | 9.69 | 0.12 | 10.68 | 22.02 | 0.40 | 0.02 | 99.18 | 0.55 | 0.08 | 0.00 | 0.07 | 0.33 | 0.84 | 49.55 | 33.43 | 17.02 |
| Cpx5_A3 | 43.93 | 2.17 | 9.90 | 8.84 | 0.06 | 11.72 | 22.41 | 0.33 | 0.00 | 99.36 | 0.57 | 0.09 | 0.00 | 0.08 | 0.29 | 0.86 | 49.13 | 35.75 | 15.13 |
| Cpx5_A3 | 42.52 | 2.50 | 10.68 | 9.48 | 0.21 | 10.86 | 21.91 | 0.42 | 0.01 | 98.58 | 0.54 | 0.10 | 0.00 | 0.08 | 0.31 | 0.83 | 49.33 | 34.01 | 16.66 |
| Cpx5_A3 | 43.43 | 2.14 | 11.09 | 9.18 | 0.16 | 10.90 | 21.69 | 0.52 | 0.01 | 98.88 | 0.52 | 0.10 | 0.00 | 0.09 | 0.31 | 0.83 | 49.26 | 34.45 | 16.28 |

| Plagioclase | SiO2 | TiO2 | Al2O3 | FeO | MnO | MgO | CaO | Na2O | K2O | Total | An% | Ab% | Or% |
|---|---|---|---|---|---|---|---|---|---|---|---|---|---|
| Plg1_Sec2_A1 | 52.14 | 0.11 | 29.00 | 0.99 | 0.03 | 0.13 | 12.14 | 4.18 | 0.38 | 99.11 | 60.23 | 37.53 | 2.24 |
| Plg1_Sec2_A1 | 52.45 | 0.10 | 29.01 | 1.12 | 0.01 | 0.12 | 12.16 | 4.32 | 0.37 | 99.68 | 59.57 | 38.26 | 2.17 |
| Plg1_Sec2_A1 | 51.86 | 0.10 | 29.02 | 1.09 | 0.03 | 0.14 | 12.17 | 4.19 | 0.38 | 98.99 | 60.26 | 37.52 | 2.22 |
| Plg3_Sec2_A1 | 53.18 | 0.16 | 28.36 | 1.14 | 0.04 | 0.11 | 11.29 | 4.79 | 0.42 | 99.55 | 54.69 | 33.71 | 1.60 |
| Plg3_Sec2_A1 | 51.56 | 0.06 | 29.65 | 1.08 | -0.03 | 0.18 | 13.03 | 3.75 | 0.27 | 99.55 | 61.43 | 33.73 | 1.47 |
| Plg4_Sec2_A2 | 52.06 | 0.08 | 29.28 | 1.25 | -0.03 | 0.14 | 12.29 | 4.05 | 0.33 | 99.45 | 61.43 | 36.60 | 1.97 |

| Oxide | SiO2 | TiO2 | Al2O3 | FeO | MnO | MgO | CaO | Na2O | K2O | Total |
|---|---|---|---|---|---|---|---|---|---|---|
| Oxide | 0.09 | 5.64 | 10.81 | 69.68 | 0.40 | 8.47 | 0.19 | 0.02 | 0.02 | 95.32 |
| Oxide | 0.09 | 6.58 | 7.93 | 71.48 | 0.48 | 8.03 | 0.11 | 0.02 | 0.01 | 94.61 |
| Oxide | 0.10 | 6.31 | 8.16 | 72.19 | 0.41 | 8.11 | 0.04 | 0.02 | 0.01 | 95.28 |
| Oxide | 0.12 | 6.33 | 8.22 | 72.60 | 0.58 | 8.09 | 0.13 | 0.02 | 0.03 | 96.19 |

### 100_S4
Experimental conditions:  0.9 XCO2; P=100 Mpa; T 1100°C

| spots | SiO2 | TiO2 | Al2O3 | FeO | MnO | MgO | CaO | Na2O | K2O | Total | Diopside | Hedenbergit | Jadeite | Opx | Tschermakite | Mg# | Wo | En | Fs |
|---|---|---|---|---|---|---|---|---|---|---|---|---|---|---|---|---|---|---|---|
| Cpx1_A1 | 43.05 | 2.33 | 11.02 | 9.40 | 0.18 | 11.28 | 21.06 | 0.48 | 0.02 | 98.82 | 0.58 | 0.10 | 0.00 | 0.11 | 0.31 | 0.83 | 47.76 | 35.59 | 16.64 |
| Cpx2_A1 | 42.60 | 2.52 | 11.28 | 8.93 | 0.15 | 10.92 | 21.10 | 0.50 | 0.01 | 98.02 | 0.59 | 0.08 | 0.00 | 0.09 | 0.33 | 0.85 | 48.77 | 35.12 | 16.11 |
| Cpx3_A1 | 41.84 | 2.72 | 11.27 | 9.63 | 0.18 | 10.74 | 21.31 | 0.51 | 0.01 | 98.16 | 0.51 | 0.09 | 0.00 | 0.09 | 0.33 | 0.85 | 48.70 | 34.13 | 17.18 |
| Cpx4_A1 | 43.22 | 2.17 | 11.18 | 9.33 | 0.12 | 11.07 | 21.52 | 0.48 | 0.01 | 99.13 | 0.51 | 0.10 | 0.00 | 0.10 | 0.31 | 0.83 | 48.69 | 34.84 | 16.47 |
|  |  |  |  |  |  |  |  |  |  |  |  |  |  |  |  | 0.83 |  |  |  |

| Plagioclase | SiO2 | TiO2 | Al2O3 | FeO | MnO | MgO | CaO | Na2O | K2O | Total | An% | Ab% | Or% |
|---|---|---|---|---|---|---|---|---|---|---|---|---|---|
| Plg1_Exp2_A1 | 51.96 | 0.15 | 29.13 | 1.12 | 0.03 | 0.12 | 11.90 | 4.29 | 0.44 | 99.13 | 58.95 | 38.49 | 2.57 |
| Plg2_Exp2_A1 | 51.46 | 0.13 | 29.28 | 1.13 | 0.02 | 0.09 | 12.09 | 4.14 | 0.43 | 99.33 | 59.97 | 37.47 | 2.55 |
| Plg2_Exp2_A1 | 52.08 | 0.11 | 29.10 | 1.10 | 0.00 | 0.10 | 12.06 | 4.28 | 0.43 | 99.36 | 59.35 | 38.14 | 2.52 |
| Plg4_Exp2_A1 | 53.75 | 0.15 | 28.32 | 1.11 | 0.01 | 0.11 | 10.66 | 4.92 | 0.56 | 99.68 | 52.70 | 44.01 | 3.29 |
| Plg5_Exp2_A1 | 53.14 | 0.11 | 28.44 | 1.10 | -0.06 | 0.12 | 11.88 | 4.63 | 0.52 | 99.88 | 55.20 | 41.73 | 3.07 |
| Plg6_Exp2_A1 | 51.07 | 0.10 | 30.13 | 0.92 | 0.09 | 0.09 | 13.08 | 3.77 | 0.35 | 99.60 | 64.36 | 33.61 | 2.03 |

| Oxides | SiO2 | TiO2 | Al2O3 | FeO | MnO | MgO | CaO | Na2O | K2O | Total |
|---|---|---|---|---|---|---|---|---|---|---|
| Oxide | 0.04 | 8.61 | 7.37 | 70.24 | 0.57 | 6.39 | 0.20 | 0.04 | 0.03 | 94.62 |
| Oxide | 0.12 | 8.49 | 7.73 | 71.28 | 0.42 | 6.71 | 0.21 | 0.01 | 0.00 | 96.20 |
| Oxide | 0.01 | 8.52 | 7.51 | 70.55 | 0.43 | 6.82 | 0.24 | 0.01 | 0.03 | 95.25 |
| Oxide | 0.11 | 8.20 | 7.95 | 70.68 | 0.44 | 6.91 | 0.19 | 0.01 | 0.03 | 95.68 |